\documentclass{article}

\usepackage[main, final]{neurips_2026}

\usepackage[utf8]{inputenc} 
\usepackage[T1]{fontenc}    
\usepackage{hyperref}       
\hypersetup{
    colorlinks=true,
    urlcolor=blue,
    citecolor=black,
    linkcolor=black
}
\usepackage{url}            
\usepackage{booktabs}       
\usepackage{amsfonts}       
\usepackage{nicefrac}       
\usepackage{microtype}      
\usepackage[table]{xcolor}  
\usepackage{graphicx}
\usepackage{wrapfig}
\usepackage{multirow}
\usepackage{array}
\usepackage{amsmath}
\usepackage{amssymb}
\usepackage{longtable}
\usepackage{subcaption}
\usepackage{pifont}
\usepackage{etoc}           
\usepackage{placeins}       

\definecolor{MyLightGreen}{RGB}{0,150,80}
\definecolor{MyLightRed}{RGB}{200,40,40}

\newcommand{\cmark}{\textcolor{MyLightGreen}{\ding{51}}}
\newcommand{\xmark}{\textcolor{MyLightRed}{\ding{55}}}

\usepackage{caption}

\title{%
  \raisebox{-1em}[0pt][0pt]{%
    \includegraphics[height=2.5em]{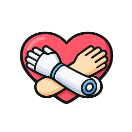}%
  }%
  \hspace{0.2em}%
  EchoChat: Structured Cognitive Reasoning \\ in Empathetic Spoken Dialogue%
}

\author{
\begin{tabular}{c}
\textbf{
Dingdong Wang\textsuperscript{1,2}\thanks{Work done during an internship at Microsoft},
Shujie Liu\textsuperscript{2}\footnotemark[2],
Yayue Deng\textsuperscript{1},
Yuxuan Hu\textsuperscript{2},
} \\
\textbf{
Yunrui Cai\textsuperscript{1},
Jincenzi Wu\textsuperscript{1},
Jianwei Yu\textsuperscript{2},
Jinyu Li\textsuperscript{2},
Helen Meng\textsuperscript{1}\footnotemark[2]\thanks{\ \ Corresponding Author}
} \\
\normalfont\textsuperscript{1}The Chinese University of Hong Kong, China \\
\normalfont\textsuperscript{2}Microsoft Corporation \\
\normalfont\texttt{dingdongwang@link.cuhk.edu.hk}
\end{tabular}
}

\begin{document}

\maketitle

\begin{abstract}

Empathetic spoken dialogue is a sophisticated cognitive process that requires not only recognizing emotions but also inferring a user’s latent mental states to provide appropriate support. However, current SpeechLLMs often treat empathy as a direct input-to-response mapping, leading to "superficially warm" but emotionally hollow interactions. In addition, since empathy relies on a multi-stage process with strong inter-step dependency, errors at any intermediate step can cascade through subsequent steps and lead to inappropriate responses, while existing training paradigms lack mechanisms to precisely localize and improve such errors. In this work, we propose EchoChat, a unified framework that reformulates empathetic spoken dialogue as a structured cognitive reasoning process integrating perception, mental-state reasoning, and response generation. To support this paradigm, we first construct EchoDialogue-400K, an acoustically rich dataset for multi-stage empathetic supervision. During the SFT stage, we strengthen acoustic grounding through proposed Acoustic-Anchored Attention (AAA). During the RL stage, we further introduce a novel stage-aware optimization objective with Step-Decomposed Credit Assignment (SDCA) to localize reasoning errors and mitigate cascaded error propagation. In addition, we introduce EchoEval, an expert-annotated benchmark for multi-dimensional empathy evaluation. Extensive experiments demonstrate that EchoChat achieves state-of-the-art performance in perception, reasoning, and response alignment. The \href{https://github.com/dingdongwang/EchoChat}{project page},
\href{https://huggingface.co/ddwang2000/EchoChat}{EchoChat model},
\href{https://huggingface.co/datasets/ddwang2000/EchoEval}{benchmark},
and \href{https://huggingface.co/datasets/ddwang2000/EchoDialogue}{EchoDialogue dataset}
are publicly available and open-sourced.
\end{abstract}

\etocdepthtag{paper}
\section{Introduction}

Empathy is the cornerstone of human-centric interaction. It requires a system to understand not only what a user says, but also how the user feels and what kind of support the user actually needs~\cite{cuff2016empathy, rashkin2019towards}. While recent advances in Speech Large Language Models (SpeechLLMs) have achieved remarkable fluency~\cite{arora2025landscape,ji2024wavchat,xu2025qwen25omnitechnicalreport,Qwen2-Audio}, they often fall into the trap of "empathetic mimicry"~\cite{hess2014emotional}, producing responses that are superficially warm but fundamentally misaligned with the user's latent mental state. This stems from a core limitation: existing systems treat empathy as a direct input-to-response generation problem~\cite{wang2025opens2s,wang2026covo,yang2025paras2s,geng2025osum}, essentially bypassing the structured cognitive reasoning process. In reality, empathetic dialogue requires perceiving fine-grained acoustic and semantic cues, inferring the user's latent mental state and underlying needs, and planning an appropriate response before speaking. This challenge is particularly pronounced in the speech domain, where subtle prosodic variations can substantially reshape the emotional context.


The simplified, response-centric formulation of empathetic spoken dialogue introduces three fundamental limitations. (1) \textbf{Inadequate affective perception.} Existing studies model user affect using coarse-grained emotion categories (e.g., "happy", "angry")~\cite{wang2025opens2s,geng2025osum,jia2026reflecting,wang2025empathy,hu2025chain}, failing to capture the subtle acoustic variations essential for authentic empathy. For example, within the same sadness category, intense sorrow accompanied by sobbing requires a fundamentally different support strategy from mild melancholy. This issue becomes more pronounced in complex scenarios, such as emotional shifts within an utterance and sarcasm, which current models find difficult to handle due to their limited perceptual granularity. Yet, existing datasets~\cite{geng2025osum,wang2025opens2s} provide limited coverage of such fine-grained acoustic variations, restricting models from learning the broad spectrum of human affect. (2) \textbf{Vulnerability to cascaded error propagation.} Since empathetic reasoning involves strong dependency across intermediate steps, errors in early steps can propagate through subsequent steps and lead to inappropriate responses. Although recent works introduce Chain-of-Thought (CoT) as intermediate analysis~\cite{jia2026reflecting,xie2025leveraging}, they lack robust stage-wise credit assignment. As a result, such errors become difficult to localize and improve. (3) \textbf{Lack of unified optimization across empathetic stages.} Prior approaches often optimize individual stages in isolation~\cite{geng2025osum,wang2025opens2s,wang2026covo,wang2025empathy}, such as emotion recognition and speech generation. However, empathetic spoken dialogue requires coordinated alignment across perception, reasoning, and response generation. Without a unified objective, gains at one stage may not reliably transfer to subsequent stages. ParaS2S~\cite{yang2025paras2s} takes a step further by exploring perception-to-speech-response alignment, but still bypasses the intermediate reasoning that explains \emph{what} the user actually needs and \emph{why} a specific support strategy is appropriate.

\begin{figure*}[!t]
    \centering
\includegraphics[width=1\linewidth]{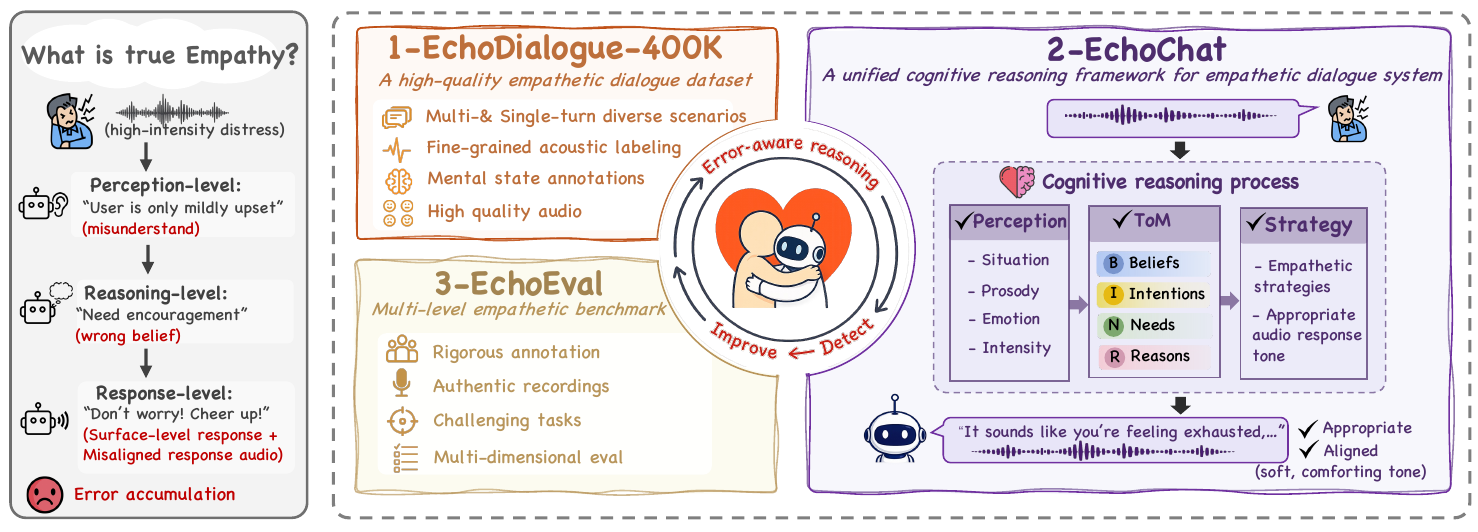}
    \caption{\small Reframing empathetic dialogue as structured cognitive reasoning. Existing methods simplify empathy to direct response generation, leading to cascaded reasoning errors across intermediate stages. EchoChat instead models empathy as a structured process from perception to response, supported by EchoDialogue-400K and EchoEval, enabling cognitively grounded and emotionally aligned spoken interaction.}
    \label{fig:exp-overview}
    \vspace{-1mm}
\end{figure*}

To address these limitations, we introduce \textbf{EchoChat}, a unified framework that models empathetic spoken dialogue as a structured process of perception, latent mental-state reasoning, and response generation. The mental-state reasoning is grounded in Theory-of-Mind (ToM) principles~\cite{frith2005theory}, which model the ability to infer others’ beliefs, intentions, needs, and psychological states from observable signals~\cite{frith2005theory}. First, to support fine-grained perception, we construct EchoDialogue-400K, a high-quality empathetic dialogue dataset covering diverse acoustic expressions and complex affective scenarios. We further introduce Acoustic-Anchored Attention (AAA) during supervised fine-tuning to strengthen the grounding of perception in speech signals. Second, to address cascaded error propagation, we design a stage-aware cascaded reasoning reward with Step-Decomposed Credit Assignment (SDCA). The cascaded reward evaluates perception, mental-state analysis, and strategy planning in a dependency-aware manner. Moreover, unlike vanilla GRPO~\cite{guo2025deepseek}, which broadcasts a single advantage across all generated tokens, SDCA decomposes stage-level signals into token-level credit, allowing errors to be localized and corrected across the reasoning trajectory. Finally, we combine the reasoning reward with a multimodal response reward under a unified training framework to jointly optimize perception, reasoning, and emotionally aligned speech generation.

For evaluation, current empathetic dialogue benchmarks~\cite{zhan2026vstyle,yan2025uro,zhou2025echomind,geng2025osum,yang2025paras2s} are limited in both stage coverage and scenario complexity, preventing comprehensive assessment of empathetic spoken dialogue. To address this, we propose EchoEval, a challenging expert-annotated benchmark derived from authentic recordings to evaluate empathy across perception, reasoning, and response generation. EchoEval encompasses diverse real-world scenarios such as sarcasm, varying emotional intensities, and dynamic emotion shifts. Extensive experiments show that EchoChat consistently improves perception, reasoning, and response alignment over strong baselines, validating the effectiveness of modeling empathy as a structured cognitive reasoning process.

\etocdepthtag{paper}

\etocdepthtag{paper}
\section{EchoDialogue-400K}

\begin{table*}[t]
\centering
\scriptsize
\setlength{\tabcolsep}{1.8pt}
\renewcommand{\arraystretch}{1.05}
\caption{\small Comparison with existing speech empathetic dialogue datasets. EchoDialogue-400K supports structured empathetic reasoning with fine-grained affective annotations and rich acoustic variations. }
\label{tab:dataset_comparison}
\resizebox{\textwidth}{!}{
\begin{tabular}{lccccccccc}
\toprule
\textbf{Dataset} 
& \textbf{Scale} 
& \textbf{Open Source}
& \textbf{Multi-Turn}
& \textbf{Reason.} 
& \textbf{Mental State} 
& \textbf{Prosody Labels} 
& \textbf{Intensity} 
& \textbf{Sarcasm} 
& \textbf{Affective Granularity} \\
\midrule
EChat-200K~\cite{geng2025osum}
& 200K 
& \xmark
& \xmark 
& \xmark 
& \xmark 
& \xmark 
& \xmark 
& \xmark 
& Coarse emotion labels \\

EmotionalQA~\cite{wang2025empathy}
& 200K 
& \xmark
& \xmark 
& \xmark 
& \xmark 
& \xmark 
& \xmark 
& \xmark 
& Coarse emotion labels \\

ParaS2S~\cite{yang2025paras2s}
& 100K
& \xmark
& \xmark 
& \xmark 
& \xmark 
& \xmark 
& \xmark 
& \cmark 
& Coarse emotion labels \\

OpenS2S~\cite{wang2025opens2s} 
& 100K
& \cmark
& \xmark 
& \cmark 
& \xmark 
& \xmark 
& \xmark 
& \xmark 
& Coarse emotion labels \\

\textbf{EchoDialogue-400K} 
& \textbf{400K} 
& \textbf{\cmark}
& \cmark 
& \cmark 
& \cmark 
& \cmark 
& \cmark 
& \cmark 
& \textbf{Fine-grained variants} \\
\bottomrule
\end{tabular}
}
\vspace{-2mm}
\end{table*}

\begin{figure*}[!t]
    \centering
\includegraphics[width=1\linewidth]{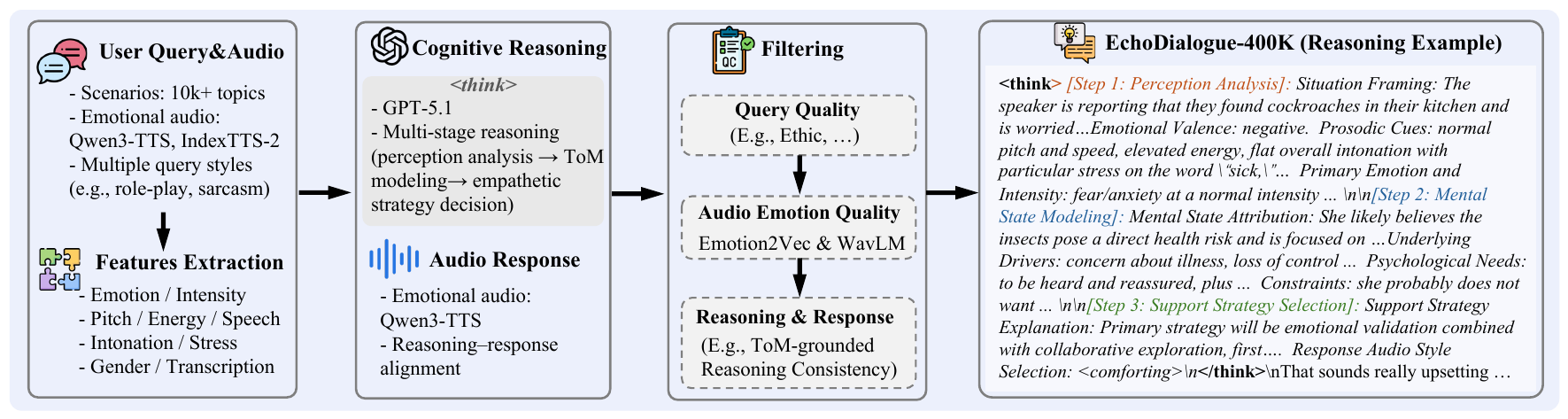}
    \caption{\small Data construction pipeline for EchoDialogue-400K.}
    \label{fig:data-pipeline}
    \vspace{-1mm}
\end{figure*}

\label{sec:EchoDialogue}
As shown in Table~\ref{tab:dataset_comparison}, prior empathetic dialogue datasets offer limited fine-grained acoustic variation (e.g., prosody) and user mental-state annotation, which hinders grounded perception and structured reasoning. To address this, we introduce EchoDialogue-400K, which supports joint learning of perception, Theory-of-Mind (ToM)~\cite{frith2005theory} reasoning, and speech generation. It contains 400K single-turn samples and 4K multi-turn dialogue sessions comprising 20K turns. To our knowledge, this is the first dataset providing integrated cognitive and acoustic supervision for spoken empathetic dialogue.

To address the limited scenario diversity and lack of expressive variation in prior corpora, we construct a structured taxonomy comprising 5 domains, 17 subcategories, and 10K+ topics (Figure~\ref{fig:EchoDialogue-structure}). The emotion space covers six basic categories (happy, sad, angry, surprise, disgust, and fear) with fine-grained variants (e.g., anxiety within fear) and varying intensities. Figure~\ref{fig:data-pipeline} outlines our data generation and filtering pipeline: we sample user queries from multiple LLMs under varied prompts (neutral, sarcastic, role-play) on predefined scenarios; audio is synthesized using diverse TTS strategies combined with rigorous filtering to ensure expressive and accurate emotional delivery. Structured reasoning traces are generated to maintain consistency in ToM inference and strategy choice. Details on data statistics, construction, and filtering are provided in the Appendix~\ref{sec:append-EchoDialogue}.

\section{Method}

\begin{figure*}[!t]
    \centering
\includegraphics[width=1\linewidth]{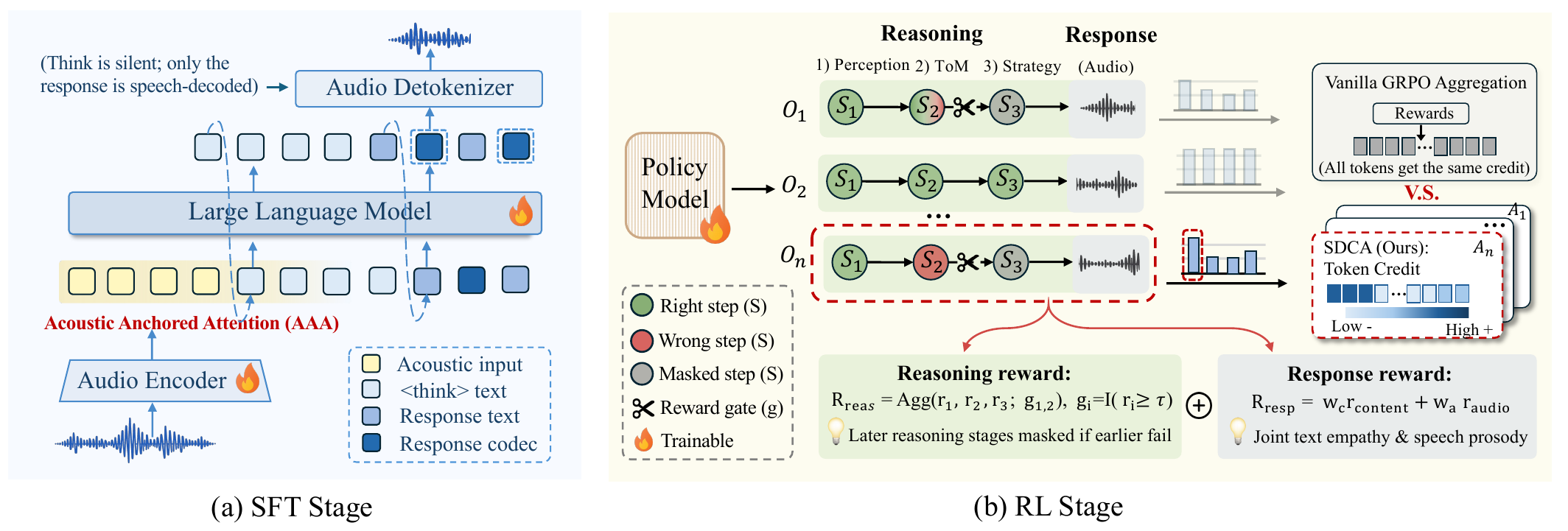}
    \caption{\small Overview of the EchoChat framework. (a) SFT uses Acoustic-Anchored Attention (AAA) to ground reasoning in acoustic signals. (b) RL performs structured reasoning optimization, where Step-Decomposed Credit Assignment (SDCA) and gating mitigate error propagation, and a response reward model aligns content and speech. Unlike vanilla GRPO, credit is stage-aware at the token level rather than uniform.}
    \label{fig:exp-arch}
    \vspace{-1mm}
\end{figure*}

We formulate empathetic spoken dialogue as a structured process that jointly models acoustic perception, Theory-of-Mind reasoning, and response generation. Given user speech, the model generates a reasoning trace (\texttt{<think>}) and an emotionally aligned response, ensuring generation is grounded in inferred mental states rather than surface patterns. Figure~\ref{fig:exp-arch} summarizes EchoChat: Stage~1 is acoustic-grounded SFT with Acoustic-Anchored Attention (AAA) (Sec.~\ref{sec:aaa}); Stage~2 is structured RL (Sec.~\ref{sec:structured-rl}) that jointly optimizes perception, reasoning and responses via cascaded reasoning reward, response-level reward and Step-Decomposed Credit Assignment (SDCA).

\subsection{Supervised Fine-Tuning (SFT)}
\label{sec:aaa}

We observe a common failure mode in chain-of-thought (CoT)  reasoning for SpeechLLMs~\cite{wei2022chain,lyu2023faithful,xu2025qwen3,tian2025step}: the model may produce spurious reasoning that mimics the surface form of explanations and follows next-token prediction patterns, rather than consistently conditioning on the input audio. However, standard supervised fine-tuning does not enforce a structural dependency between intermediate reasoning and the audio stream. As the reasoning trace grows, later steps become less grounded in acoustic cues, leading to misinterpretation of the user’s emotional state.

To address this, we introduce \textbf{Acoustic-Anchored Attention (AAA)}, a lightweight regularization mechanism that explicitly encourages perception-stage reasoning to attend to acoustic representations. Given the structured reasoning trace, we identify perception-related tokens $\mathcal{P}$ (perception stage tokens inside the structured reasoning span delimited by \texttt{<think>}) and acoustic tokens $\mathcal{A}$,  where the latter correspond to positions whose keys and values are derived from the speech encoder. For attention weights $\alpha_{i,j}$, we measure the average attention mass allocated from perception tokens to acoustic tokens as
\begin{equation}
\rho =
\frac{1}{|\mathcal P|}
\sum_{i\in \mathcal P}
\sum_{j\in \mathcal A}
\alpha_{i,j},
\end{equation}
and maximize this quantity via an auxiliary objective
$\mathcal{L}_{\text{AAA}} = -\rho$.

The overall training objective is
\begin{equation}
\mathcal{L}_{\text{SFT}} = \mathcal{L}_{\text{CE}} + \lambda_{\text{AAA}} \cdot \mathcal{L}_{\text{AAA}}.
\end{equation}
AAA imposes a soft structural constraint by linking perception-stage queries to acoustic evidence through attention allocation, without modifying model architecture or requiring additional supervision. This improves acoustic grounding during SFT and provides a stronger foundation for subsequent reasoning and response optimization as validated in our experiments.

\subsection{Structured Reinforcement Learning (RL)}
\label{sec:structured-rl}

While Acoustic-Anchored Attention improves grounding during SFT, it remains insufficient for multi-step reasoning. In CoT reasoning, errors at intermediate steps can propagate, as subsequent reasoning is built on potentially flawed earlier assumptions. Standard GRPO training assigns rewards only at the trajectory level, making it difficult to precisely localize and correct such errors. Moreover, empathetic dialogue requires unified optimization across perception, reasoning, and response, which cannot be achieved by independent or response-level objectives. To address these limitations, we propose a structured reinforcement learning framework that defines two complementary reward dimensions—a \emph{reasoning-level} reward with cascaded gating (Sec.~\ref{sec:cascade}) and a \emph{response-level} reward from a multimodal reward model (Sec.~\ref{sec:response-rm})—and introduces Step-Decomposed Credit Assignment (Sec.~\ref{sec:sdca}) to convert these rewards into stage-aware, per-token advantages.

\subsubsection{Reasoning-Level Reward with Cascaded Gating}
\label{sec:cascade}

The key insight of cascaded reward gating is that later reasoning should only be rewarded when earlier reasoning is reliable. In empathetic spoken dialogue, an appropriate support strategy depends on a plausible mental-state inference, which in turn depends on grounded and accurate emotion perception. Therefore, we evaluate the reasoning trace as a staged process rather than as an unordered block of text. Specifically, for each rollout $i$, a frozen LLM-as-judge assigns three rubric-aligned scores $s^{(i)}_1, s^{(i)}_2, s^{(i)}_3 \in [0,5]$ to the perception, mental-state inference, and support-strategy stages, respectively, which we normalize as $\tilde{s}^{(i)}_k = s^{(i)}_k / 5 \in [0,1]$. To model stage dependency, we define hard gates on the normalized scores:
\begin{equation}
  g^{(i)}_1 = \mathbb{I}\!\left(\tilde{s}^{(i)}_1 \ge \tau\right), \qquad
  g^{(i)}_2 = \mathbb{I}\!\left(\tilde{s}^{(i)}_2 \ge \tau\right),
\end{equation}
where $\tau \in [0,1]$ is a stage-quality threshold. The cascaded reasoning reward is then defined as
\begin{equation}
  R^{(i)}_{\mathrm{reas}}
  =
  \alpha_1 \tilde{s}^{(i)}_1
  +
  \alpha_2 \tilde{s}^{(i)}_2 g^{(i)}_1
  +
  \alpha_3 \tilde{s}^{(i)}_3 g^{(i)}_1 g^{(i)}_2,
  \qquad
  \sum_{k=1}^{3} \alpha_k = 1,\;\; \alpha_k \ge 0.
  \label{eq:cascade-reward}
\end{equation}
This design suppresses shortcut trajectories: if perception is unreliable, downstream inference and strategy scores do not contribute to the reasoning reward, even if their surface form appears plausible. This enforces a causal dependency in reward assignment, preventing later stages from compensating for errors in earlier reasoning.

\subsubsection{Response-Level Reward and Reward Model}
\label{sec:response-rm}

Beyond internal reasoning, empathetic dialogue requires a high-quality speech response. We train a response-level reward model based on Qwen2.5-Omni~\cite{xu2025qwen25omnitechnicalreport} to evaluate whether the final response is both semantically empathetic and acoustically aligned with the inferred emotional state. To construct training data, we generate paired positive and negative samples that vary in content quality and audio–emotion alignment. Specifically, we sample 50k instances from EchoDialogue as positive samples, and use prompt-based augmentation to produce responses with different levels of content quality. In addition, we synthesize speech with diverse emotional styles using Qwen3-TTS~\cite{hu2026qwen3} to create acoustically mismatched negative samples. This design enables supervision of both content-level empathy and speech-level emotional alignment.

The reward model takes as input the user audio, inferred emotional state, and generated speech response, and predicts two complementary signals: a content-level empathy score $s_{\mathrm{content}} \in [0,5]$, and an audio-alignment score $r_{\mathrm{audio}} \in \{0,1\}$. The content score evaluates whether the response is supportive, contextually appropriate, and coherent, while the audio-alignment score measures whether the emotional expression matches the user's state. Details of data construction and training are provided in the Appendix~\ref{sec:append-rewardmodel}.

We normalize the content score to $[0,1]$ via $r_{\mathrm{content}} = s_{\mathrm{content}} / 5$, and combine the two signals to form the response-level reward:
\begin{equation}
  r^{(i)}_{\mathrm{resp}} =
  \omega_{\mathrm{c}} \, r^{(i)}_{\mathrm{content}} +
  \omega_{\mathrm{a}} \, r^{(i)}_{\mathrm{audio}},
  \qquad
  \omega_{\mathrm{c}} + \omega_{\mathrm{a}} = 1.
  \label{eq:response-reward}
\end{equation}
We additionally compute a binary format reward $r^{(i)}_{\mathrm{fmt}} \in \{0,1\}$ that verifies compliance with the required output structure (delimited reasoning span and stage markers). Together, $r^{(i)}_{\mathrm{resp}}$ and $r^{(i)}_{\mathrm{fmt}}$ constitute the response-level reward channel.

\subsubsection{Step-Decomposed Credit Assignment (SDCA)}
\label{sec:sdca}

The reasoning-level reward $R^{(i)}_{\mathrm{reas}}$ and the response-level signals $(r^{(i)}_{\mathrm{resp}}, r^{(i)}_{\mathrm{fmt}})$ are combined into a single trajectory-level scalar
\begin{equation}
  R^{(i)}_{\mathrm{total}}
  =
  w_{\mathrm{reas}}\, R^{(i)}_{\mathrm{reas}}
  +
  w_{\mathrm{resp}}\, r^{(i)}_{\mathrm{resp}}
  +
  w_{\mathrm{fmt}}\, r^{(i)}_{\mathrm{fmt}},
  \qquad
  w_{\mathrm{reas}} + w_{\mathrm{resp}} + w_{\mathrm{fmt}} = 1.
  \label{eq:total-reward}
\end{equation}
Standard GRPO would compute a single scalar advantage from $R^{(i)}_{\mathrm{total}}$ and broadcast it to every generated token. This is coarse for structured reasoning: tokens belonging to the perception, inference, strategy, and response spans differ in their causal contribution to the reward and should not receive identical credit. We therefore introduce \emph{Step-Decomposed Credit Assignment} (SDCA), which decomposes credit across the four spans of the trajectory.

Let $\mathcal{G}(p)$ denote the set of rollouts sampled for prompt $p$. SDCA defines a per-span scalar reward for each rollout: for the three reasoning spans we directly reuse the gated stage components of $R^{(i)}_{\mathrm{reas}}$ in Eq.~\ref{eq:cascade-reward}, denoted $r^{(i)}_k$ for $k\in\{1,2,3\}$; for the response span we additionally form
\begin{equation}
  r^{(i)}_{\mathrm{R}} = \tilde\omega_{\mathrm{resp}}\, r^{(i)}_{\mathrm{resp}} + \tilde\omega_{\mathrm{fmt}}\, r^{(i)}_{\mathrm{fmt}},
  \qquad
  \tilde\omega_{\mathrm{resp}} + \tilde\omega_{\mathrm{fmt}} = 1,\;\; \tilde\omega_{\mathrm{resp}}, \tilde\omega_{\mathrm{fmt}} \ge 0.
\end{equation}
For each span $k \in \{1,2,3,\mathrm{R}\}$ we compute an independent group-relative advantage,
\begin{equation}
  A^{(i)}_k
  =
  \frac{r^{(i)}_k - \mu_k(p)}{\sigma_k(p) + \varepsilon},
  \qquad
  \mu_k(p)
  =
  \frac{1}{|\mathcal{G}(p)|}
  \sum_{j \in \mathcal{G}(p)} r^{(j)}_k,
  \label{eq:sdca-advantage}
\end{equation}
where $\sigma_k(p)$ is the within-group standard deviation and $\varepsilon$ is a small constant. We then partition the generated sequence into spans aligned with these four roles and assign $A^{(i)}_k$ to every token in span $k$, yielding a per-token advantage tensor in place of the broadcast scalar. If span boundaries cannot be reliably recovered for a sample (e.g., parsing failure), we fall back to the standard trajectory-level advantage derived from $R^{(i)}_{\mathrm{total}}$, which is therefore retained as both the optimization signal in the non-SDCA case and a safeguard for SDCA. 

Together, the cascaded reasoning reward, the multimodal response reward, and SDCA align the GRPO objective with the staged nature of empathetic reasoning while keeping reasoning and response supervision in a single, jointly optimized loop.

\begin{figure*}[!t]
    \centering
\includegraphics[width=1\linewidth]{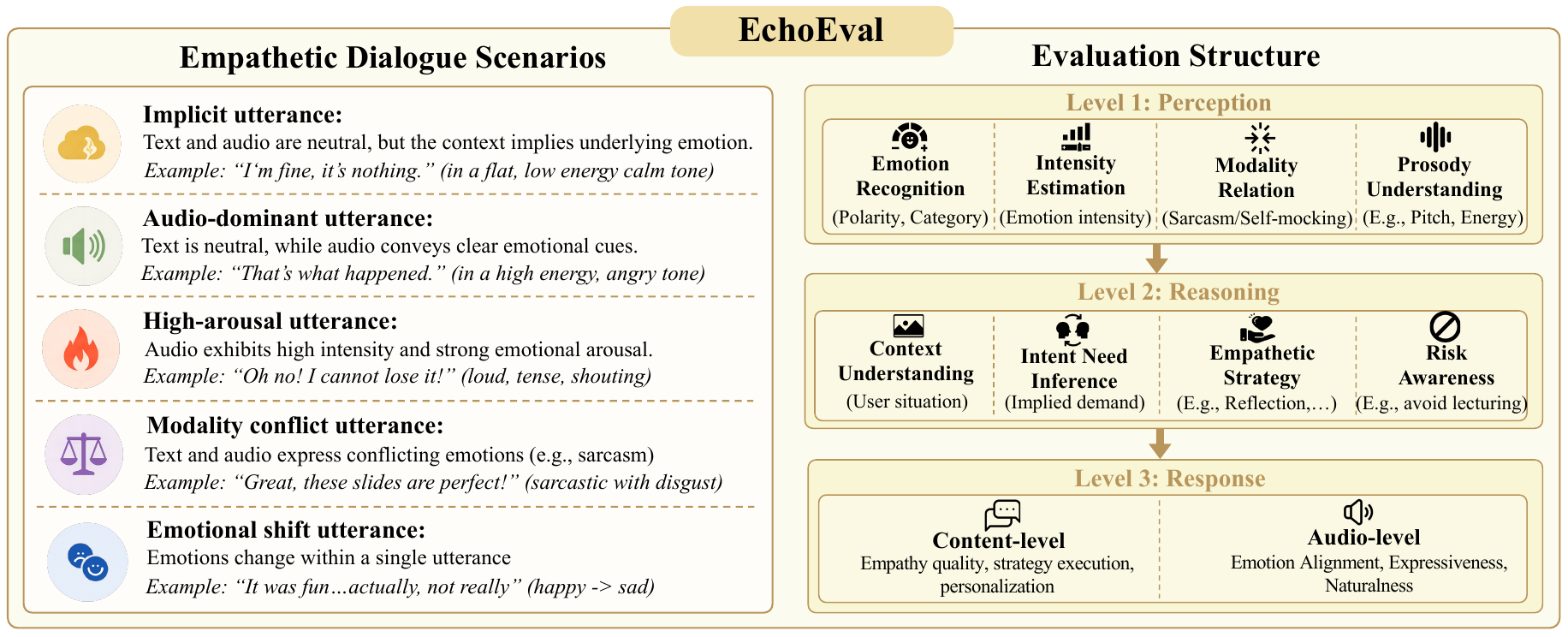}
    \caption{\small EchoEval benchmark overview. Left: diverse empathetic dialogue scenarios. Right: a three-level evaluation pipeline (perception, reasoning, response) that measures models’ empathetic ability.}
    \label{fig:EchoEval-breakdown}
    \vspace{-1mm}
\end{figure*}

\etocdepthtag{paper}
\section{EchoEval}

\paragraph{Benchmark Overview.}
Existing benchmarks for empathetic dialogue primarily focus on response-level quality~\cite{zhan2026vstyle,yan2025uro,geng2025osum,yang2025paras2s}, lacking a comprehensive and objective assessment of the full pipeline required for empathetic interaction. To address this gap, we introduce EchoEval (Figure~\ref{fig:EchoEval-breakdown}), a benchmark for systematically assessing the empathetic capabilities of SpeechLLMs, comprising 1K carefully curated instances. EchoEval is characterized by: (1) \textbf{authentic recordings} from 20 professional actors, ensuring realistic prosodic and emotional nuance; (2) \textbf{expert annotation} by psychology-trained professionals for high-quality references and rubrics; (3) \textbf{complex scenarios}, covering five challenging categories: implicit, audio-dominant, high-arousal, modality-conflict, and emotional-shift utterances; and (4) \textbf{multi-dimensional evaluation}, assessing perception, reasoning, and response through fine-grained, expert-validated criteria. Details on data curation and quality control are provided in the Appendix~\ref{sec:append-bench}.

\paragraph{Empathetic Scenarios.}


In real-world empathetic dialogue, the key challenge lies not in explicit emotion recognition but in inferring underlying mental states from subtle, implicit, or even conflicting signals. To this end, we design representative scenarios that capture complex emotional expressions in human interactions. Specifically, EchoEval covers five aspects: (1) \textbf{implicit expression}, where emotions are not explicitly stated and must be inferred from context; (2) \textbf{audio-dominant cases}, where text remains neutral while emotion is conveyed through prosody; (3) \textbf{high-arousal conditions}, requiring sensitivity to emotional intensity and appropriate response strategies; (4) \textbf{modality conflict}, such as sarcasm and self-mockery, where text and speech convey inconsistent signals and require cross-modal reasoning; and (5) \textbf{emotion shift}, where emotional states evolve within a single utterance, requiring the model to capture temporal progression. Detailed statistics are provided in the Appendix~\ref{sec:append-bench}.

\paragraph{Evaluation Setup.}

We evaluate models across three stages of empathetic dialogue: perception, reasoning, and response, with guidance from experts in psychology and cognitive science. Perception assesses recognition of context, emotion, intensity, and prosodic cues. Reasoning evaluates whether inferred latent emotional needs and corresponding response strategies are appropriate. Response examines both content quality and emotional consistency across text and speech. Specifically, each stage is evaluated using structured scoring protocols with LLM-as-judge. Perception and reasoning are assessed against reference annotations using an accumulation-based scoring scheme, where each stage receives an aggregate score on a 10-point scale. At the response stage, we adopt a two-part evaluation: content-level empathy and speech quality, each scored on a 5-point scale, assessing factors such as appropriateness, emotional tone, prosody, and naturalness. Detailed evaluation criteria, prompts, and metric definitions are provided in the Appendix~\ref{sec:append-bench}.

\etocdepthtag{paper}
\section{Experiments}
\label{sec:experiments}

\subsection{Experimental Setup}
\label{sec:exp-setup}

EchoChat is initialized from Kimi-Audio~\cite{kimiteam2025kimiaudio} and trained in two stages. In the first stage, we perform SFT on 100k samples from EchoDialogue-400K with Acoustic-Anchored Attention (AAA), where the audio encoder, adapter, and LLM are jointly optimized. In the second stage, we apply structured reinforcement learning over the reasoning--response trajectory. We adopt LoRA for efficient fine-tuning and train for 20k steps with a learning rate of $1\mathrm{e}{-5}$ and a KL divergence coefficient of 0.04. For each input, four candidate rollouts are sampled to estimate group-level reliability and compute reward signals. The overall reward is defined as a weighted sum of cascaded reasoning reward, response reward, and format reward, with weights 0.5, 0.4, and 0.1, respectively.

We primarily evaluate on EchoEval, with VStyle and human evaluation as complementary assessments. To comprehensively benchmark our approach, we compare against 12 state-of-the-art models across three categories: (1) general end-to-end SpeechLLMs, including GLM-4-Voice~\cite{zeng2024glm}, Step-Audio2 Mini~\cite{wu2025step}, Kimi-Audio~\cite{kimiteam2025kimiaudio}, Mimo-Audio-7B-Instruct~\cite{zhang2025mimo}, Covo-Audio~\cite{wang2026covo}, and Fun-Audio-Chat~\cite{team2025fun}; (2) empathy-focused SpeechLLMs, including OpenS2S~\cite{wang2025opens2s}, MERaLiON~\cite{he2025meralion}, and BLSP-Emo~\cite{wang2024blsp}; and (3) Omni Large Language Models, including MiniCPM-O-4.5~\cite{cui2026minicpm}, Qwen2.5-Omni-7B~\cite{xu2025qwen25omnitechnicalreport}, and Qwen3-Omni~\cite{xu2025qwen3}. To ensure consistent comparison across models on EchoEval, we use unified prompts for each evaluation stage inference, as detailed in the Appendix~\ref{sec:append-bench}.

\subsection{Main Results}
\label{sec:main-results}

\begin{table}[t]
    \centering
    \small  
    \caption{\small Performance comparison across models on EchoEval. Perception, reasoning, and response scores are reported on a 10-point scale; the response score consists of content quality (5 points) and audio quality (5 points). For speech-to-text models (MERaLiON and BLSP-Emo), the response score is computed using content quality only. The top two results are highlighted in \textbf{bold} and \underline{underline}, respectively.}
    \vspace{1mm}
    \renewcommand\tabcolsep{3pt}
    \resizebox{\textwidth}{!}{
    \begin{tabular}{lcccccccccccc}
    \toprule
    \textbf{Models} 
    & \multicolumn{6}{c}{\textbf{Perception}$\uparrow$} 
    & \multicolumn{1}{c}{\textbf{Reasoning}$\uparrow$} 
    & \multicolumn{3}{c}{\textbf{Response}$\uparrow$} 
    & \multicolumn{1}{c}{\textbf{All}$\uparrow$} \\
    
    \cmidrule(lr){2-7} \cmidrule(lr){8-8} \cmidrule(lr){9-11} \cmidrule(lr){12-12}
    
    & \textbf{Implicit} & \textbf{Neutral} & \textbf{Conflict} & \textbf{Intense} & \textbf{Shift} & \textbf{Avg.}
    & \textbf{Avg.}
    & \textbf{Content} & \textbf{Audio} & \textbf{Avg.}
    & \textbf{Avg.} \\
    \midrule 
    
    \multicolumn{12}{c}{\textit{General Speech Large Language Models}} \\
    \midrule
    
    GLM-4-Voice & 5.23 & 4.67 & 4.26 & 5.06 & 4.05 & 4.75 & 2.57 & 2.87 & 3.85 & 6.72 & 4.68 \\
    Step-Audio2 mini & 3.57 & 3.15 & 3.12 & 2.13 & 1.41 & 2.80 & 1.80 & 2.15 & 3.22 & 5.37 & 3.32 \\
    Kimi-Audio & 6.61 & 6.53 & 5.74 & 7.32 & 7.62 & 6.67 & 3.70 & 3.45 & 3.42 & 6.87 & 5.75 \\
    Mimo-Audio-7B-Instruct & \underline{6.98} & 5.70 & 5.62 & \underline{8.73} & \underline{8.44} & \underline{6.96} & \underline{6.61} & \underline{4.12} & 3.80 & 7.92 & 7.16 \\
    Covo-Audio & 6.79 & \underline{6.77} & 5.20 & 7.71 & 6.82 & 6.64 & 5.02 & 4.07 & 4.75 & \underline{8.82} & 6.83 \\
    Fun-Audio-Chat & 6.38 & 4.54 & 4.68 & 8.36 & 6.85 & 6.09 & 4.15 & 3.83 & 4.58 & 8.41 & 6.15 \\
    
    \midrule
    \multicolumn{12}{c}{\textit{Empathy-Focused Speech Large Language Models}} \\
    \midrule
    
    OpenS2S & 3.07 & 2.14 & 1.66 & 5.35 & 1.40 & 2.86 & 3.77 & 3.85 & 3.82 & 7.67 & 4.77 \\
    MERaLiON & 6.27 & 4.54 & 4.73 & 7.62 & 5.38 & 5.74 & 3.85 & 4.16 & - & 4.16 & 4.58 \\
    BLSP-Emo & 6.54 & 6.51 & 5.27 & 8.13 & 6.91 & 6.65 & 3.19 & 3.22 & - & 3.22 & 4.35 \\
    
    \midrule
    \multicolumn{12}{c}{\textit{Omni Large Language Models}} \\
    \midrule
    
    MiniCPM-O & 6.46 & 5.61 & 5.27 & 8.38 & 5.94 & 6.37 & 4.43 & 3.32 & 3.98 & 7.30 & 6.03 \\
    Qwen2.5-Omni & 3.82 & 2.84 & 4.21 & 6.89 & 6.43 & 4.68 & 2.60 & 3.37 & 3.89 & 7.26 & 4.85 \\
    Qwen3-Omni & 6.71 & 5.83 & \underline{5.90} & 8.64 & 7.33 & 6.84 & 5.48 & 4.00 & \underline{4.80} & 8.80 & 7.04 \\
    
    \midrule
    \rowcolor{gray!15}
    EchoChat & \textbf{8.41} & \textbf{7.79} & \textbf{7.36} & \textbf{8.87} & \textbf{8.73} & \textbf{8.23} & \textbf{8.43} & \textbf{4.61} & \underline{4.65} & \textbf{9.26} & 8.64 \\
    \bottomrule
    \end{tabular}
    }
    \label{tab:main_results}
    \vspace{-2mm}
    \end{table}

\begin{wraptable}{r}{0.27\columnwidth}
\setlength{\intextsep}{2pt}      
\setlength{\tabcolsep}{2pt}
\renewcommand{\arraystretch}{0.9}
\caption{\small Agreement between LLM-based and human judgments (Spearman's $\rho$).}
\vspace{-2mm}
\label{tab:human_alignment}
\centering
\begin{tabular}{lc}
\toprule
\textbf{Dimension} & $\rho$ \\
\midrule
Perception & 0.88 \\
Reasoning & 0.77 \\
Response & 0.83 \\
\midrule
\textbf{All (Avg.)} & \textbf{0.83} \\
\bottomrule
\end{tabular}
\end{wraptable}

\paragraph{Performance on EchoEval.}
Table~\ref{tab:main_results} reports the main results on EchoEval. EchoChat achieves the best overall performance across the full perception--reasoning--response pipeline. On perception, EchoChat consistently attains the top performance across all empathetic scenarios. The gains are particularly pronounced in neutral scenarios, where the textual content conveys neutral emotion and the model must rely on acoustic cues. In this setting, EchoChat achieves 7.79, outperforming the second-best model, Covo-Audio (6.77), by a clear margin. In the reasoning stage, EchoChat substantially improves the reasoning score, indicating a stronger ability to transform perceived emotional signals into coherent mental-state inference and appropriate support strategy selection.  A similar trend is observed at the response stage. While some baselines can produce fluent or emotionally warm replies, their responses are often less consistently aligned with the inferred user state. In contrast, EchoChat achieves strong performance at both the content level and the audio level, demonstrating stronger alignment between understanding and response. To ensure the reliability of the LLM-based evaluation, we further compare it with human judgments on a subset of EchoEval. As shown in Table~\ref{tab:human_alignment}, the LLM-based scores achieve an average Spearman's $\rho$ of 0.83 across perception, reasoning, and response, indicating strong rank-order agreement with human judgments and largely consistent pairwise model preferences.

\begin{table}[t]
\centering
\small

\begin{minipage}[t]{0.55\textwidth}
\centering
\caption{\small Performance on the English empathy set of VStyle Benchmark, with each dimension evaluated on a 5-point scale.}
\vspace{1mm}
\label{tab:vstyle-emotion}
\setlength{\tabcolsep}{3pt}
\begin{tabular}{lcccc}
\toprule
\textbf{Model} & \textbf{Anger} & \textbf{Sadness} & \textbf{Anxiety} & \textbf{Joy} \\
\midrule
Kimi-Audio & 3.59 & 3.97 & 3.65 & 3.46 \\
Step-Audio & 3.95 & 4.37 & 3.87 & 4.29 \\
Qwen2.5-Omni & 2.95 & 2.73 & 3.55 & 3.43 \\
Qwen3-Omni & \underline{4.96} & \underline{4.95} & \textbf{5.00} & \underline{4.95} \\
Fun-Audio-Chat & 3.64 & 4.10 & 2.90 & 3.69 \\
Covo-Audio-Chat & 4.86 & 4.83 & 4.97 & 4.80 \\
OpenS2S & 3.79 & 3.72 & 3.97 & 3.65 \\
\midrule
\rowcolor{gray!15}
EchoChat & \textbf{5.00} & \textbf{5.00} & \underline{4.98} & \textbf{4.96} \\
\bottomrule
\end{tabular}
\end{minipage}
\hfill
\begin{minipage}[t]{0.40\textwidth}
\centering
\caption{\small Human evaluation on multi-turn interactions. Content measures empathetic content quality, while audio evaluates emotional alignment, naturalness, and prosody.}
\vspace{1mm}
\label{tab:human_eval}
\setlength{\tabcolsep}{6pt}
\begin{tabular}{lcc}
\toprule
\textbf{Model} & \textbf{Content} & \textbf{Audio} \\
\midrule
Step-Audio & 2.10 & 3.00 \\
Fun-Audio-Chat & 3.90 & 3.80 \\
Kimi-Audio & 3.40 & 3.20 \\
Qwen3-Omni & \underline{4.50} & \underline{4.60} \\
Covo-Audio-Chat & 4.10 & 4.00 \\
\midrule
\rowcolor{gray!15}
EchoChat & \textbf{4.90} & \textbf{4.70} \\
\bottomrule
\end{tabular}
\end{minipage}

\vspace{-1mm}
\end{table}

\paragraph{Performance on VStyle.}
We further evaluate EchoChat on the VStyle~\cite{zhan2026vstyle} Empathy English Benchmark (Table~\ref{tab:vstyle-emotion}), which provides an external assessment of empathetic response quality across representative affective styles, including anger, sadness, anxiety, and joy. Compared with seven representative baselines, EchoChat achieves highly competitive performance across categories.

\paragraph{Human evaluation.}
To mitigate potential biases in LLM-based automatic evaluation, we conduct human evaluation on 120 anonymized samples (20 per model) across multi-turn dialogue interactions, covering EchoChat and five competitive baselines. Three annotators evaluate both response content and generated speech, each on a 5-point scale. At the content level, evaluation focuses on helpfulness, i.e., whether the response improves the user’s experience and effectively addresses the user’s needs. At the audio level, we adopt a MOS-style evaluation to assess naturalness, prosody, and emotional alignment with the user’s input. As shown in Table~\ref{tab:human_eval}, EchoChat demonstrates superiority across all evaluated dimensions.

\subsection{Analysis}
\label{sec:analysis}

\begin{table}[t]
\centering
\small
\vspace{-1mm}

\begin{minipage}[t]{0.62\textwidth}
\centering
\setlength{\tabcolsep}{2pt}
\renewcommand{\arraystretch}{0.8}
\caption{\small Ablation study on different training strategies on EchoEval. RL variants V1–V5 are built upon Baseline 2.}
\label{tab:ablation}
\begin{tabular}{@{}lc>{\centering\arraybackslash}p{1.0cm}>{\centering\arraybackslash}p{1.0cm}>{\centering\arraybackslash}p{0.55cm}@{}}
\toprule
\multirow{2}{*}{\textit{Variations}} 
& \multirow{2}{*}{\textit{Training Strategy}} 
& \multicolumn{1}{c}{Perc.} 
& \multicolumn{1}{c}{Reas.} 
& \multicolumn{1}{c}{Resp.} \\
\cmidrule{3-5}
& & Avg.$\uparrow$ & Avg.$\uparrow$ & Avg.$\uparrow$ \\
\midrule
Baseline 1 & SFT & 4.88 & 5.46 & 7.96 \\
Baseline 2 & SFT-AAA & 6.81 & 6.75 & 8.64 \\
\midrule
V1 & Cascaded+Form. & 8.01 & 8.26 & 8.40 \\
V2 & Cascaded(SDCA)+Form. & 8.20 & 8.44 & 8.49 \\
V3 & Format+Resp. & 7.87 & 7.30 & 9.23 \\
V4 & Cascaded+Form.+Resp. & 8.15 & 8.15 & 9.18 \\
V5 & Cascaded(SDCA)+Form.+Resp. & 8.23 & 8.43 & 9.26 \\
\bottomrule
\end{tabular}
\end{minipage}
\hfill
\begin{minipage}[t]{0.34\textwidth}
\centering
\setlength{\tabcolsep}{2pt}
\renewcommand{\arraystretch}{1.0}
\caption{\small Sensitivity analysis on acoustic perception. Results compare the SFT with and without AAA across emotion and intensity accuracy, as well as prosody and content scores.}
\vspace{2.5mm}
\label{tab:aaa_sensitivity}
\begin{tabular}{@{}lcc@{}}
\toprule
\textbf{Settings} & \textbf{w/o AAA} & \textbf{w/ AAA} \\
\midrule
Emotion (Acc) & 65\% & 77\% \\
Intensity (Acc) & 72\% & 91\% \\
Prosody (Score) & 3.27 & 3.96 \\
Content (Score) & 4.75 & 4.79 \\
\bottomrule
\end{tabular}
\end{minipage}

\vspace{-1mm}
\end{table}

\begin{wrapfigure}{r}{0.30\linewidth}
\vspace{-4mm}
\centering
\includegraphics[width=\linewidth]{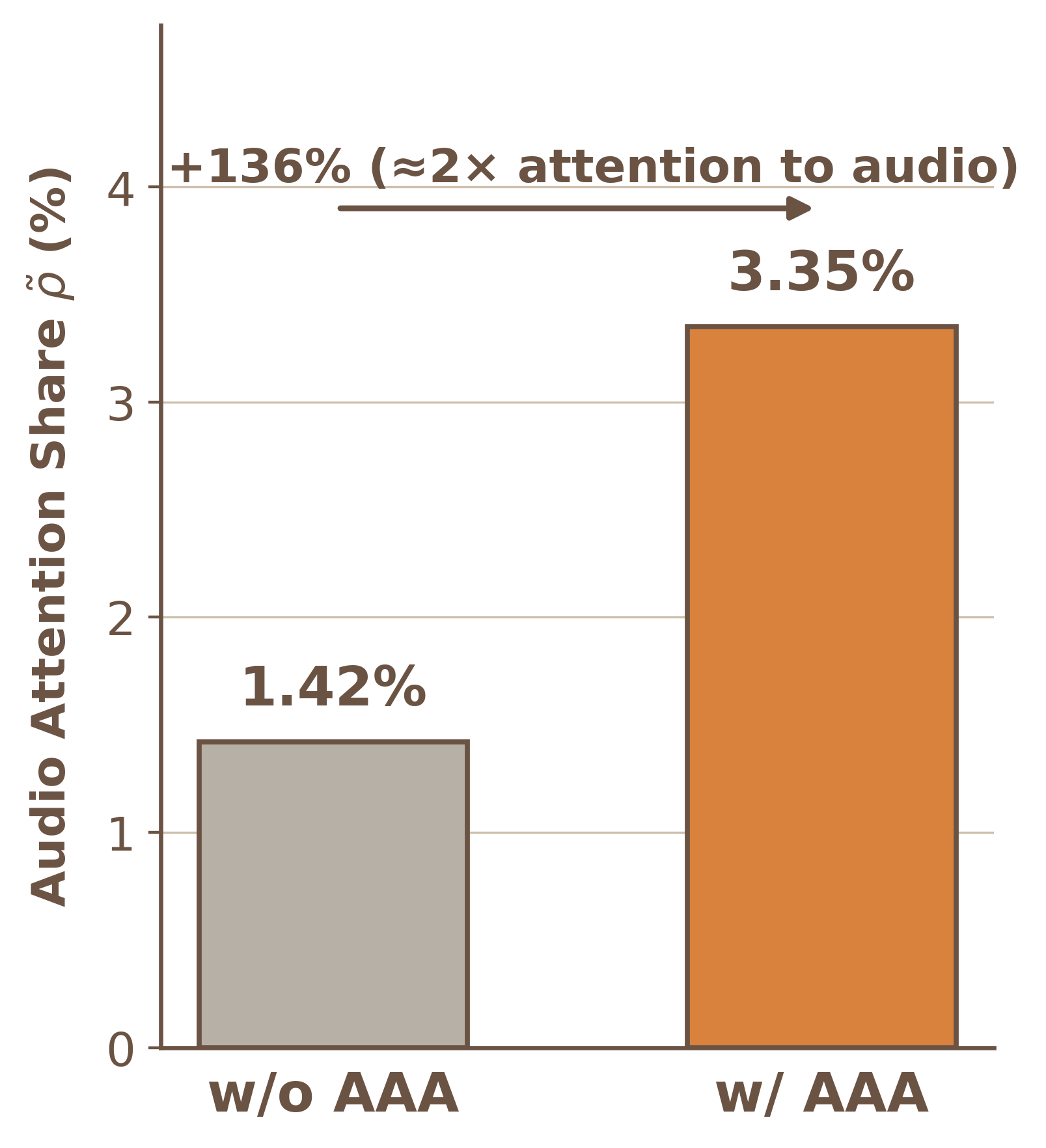}
\caption{\small AAA doubles the attention mass that perception tokens place on acoustic input $\bar{\rho}$, indicating it better grounds reasoning in audio rather than relying on textual priors.}
\label{fig:aaa_attention}
\vspace{-2mm}
\end{wrapfigure}

Table~\ref{tab:ablation} decomposes the contribution of each training signal. Plain SFT enables structured response generation, but remains limited in perception and reasoning. AAA improves both scores, showing the benefit of acoustic grounding for upstream understanding. In RL, cascaded reasoning rewards improve perception and reasoning by modeling dependencies among perception, mental-state reasoning, and strategy planning, while SDCA further localizes rewards to the corresponding reasoning spans. The response reward mainly improves response quality, and its combination with cascaded rewards and SDCA yields the best overall balance. These results highlight the benefit of jointly optimizing acoustic grounding, stage-aware reasoning, and multimodal response alignment.

We further analyze whether AAA indeed strengthens acoustic grounding. As shown in Table~\ref{tab:aaa_sensitivity}, adding AAA improves emotion and intensity accuracy, as well as prosody-related scores, indicating that the model becomes more sensitive to fine-grained acoustic cues. Figure~\ref{fig:aaa_attention} provides additional mechanism-level evidence: AAA increases the attention mass allocated to acoustic inputs during perception-stage reasoning (+136\%). This confirms that AAA encourages the model to ground its reasoning in speech signals instead of relying primarily on textual priors.


\begin{wraptable}{r}{0.44\textwidth}
\small
\setlength{\tabcolsep}{1.5pt}
\renewcommand{\arraystretch}{0.82}

\caption{\small Error propagation analysis. We report conditional correctness across stages and the overall error cascade ratio.}
\vspace{-2mm}

\begin{tabular}{@{}lccc@{}}
\toprule
\textbf{Model} & $\mathbf{P(R|P)}\uparrow$ & $\mathbf{P(Resp|R)}\uparrow$ & \textbf{Err.}$\downarrow$ \\
\midrule
Baseline & 0.52 & 0.48 & 0.41 \\
w/o gating & 0.60 & 0.51 & 0.36 \\
w/o SDCA & 0.58 & 0.55 & 0.33 \\
\textbf{EchoChat} & \textbf{0.71} & \textbf{0.68} & \textbf{0.21} \\
\bottomrule
\end{tabular}

\label{tab:error_propagation}
\vspace{-2mm}
\end{wraptable}

Table~\ref{tab:error_propagation} analyzes error propagation across the reasoning chain. $P(R|P)$ and $P(Resp|R)$ measure whether correct perception and reasoning are preserved in subsequent stages, while the error ratio measures overall cascaded failures. Gating and SDCA progressively improve conditional correctness and reduce error propagation, with EchoChat achieving the best results on all three metrics. This suggests that our structured RL design helps maintain reliable stage-wise reasoning rather than only improving final response scores.

\etocdepthtag{paper}
\section{Conclusion}

In this work, we challenge the conventional formulation of empathetic spoken dialogue as direct response generation and instead model it as a structured cognitive reasoning process. EchoChat unifies acoustic perception, latent mental-state inference, and emotionally aligned response generation within a stage-aware optimization framework, mitigating cascaded reasoning errors and improving cross-stage consistency in empathetic interaction. To support this paradigm, we further introduce EchoDialogue-400K and EchoEval for training and principled multi-stage evaluation. Extensive experiments demonstrate consistent improvements in perception, reasoning, and response alignment across diverse empathetic scenarios, highlighting the importance of modeling empathy as structured cognitive reasoning rather than relying solely on response-centric generation.

\newpage
\bibliographystyle{plain}
\bibliography{neurips_2026}

\clearpage
\FloatBarrier
\appendix

\hypersetup{
    linkcolor=black,
}

\begin{center}
 \LARGE \bf {EchoChat: Structured Cognitive Reasoning in Empathetic Spoken Dialogue\ \\ \ \\ \textit{Supplementary Material}}
\end{center}

\etocsettagdepth{paper}{none}
\etocsettagdepth{supp}{subsection}
\tableofcontents

\hypersetup{
    linkcolor=black,
}

\etocdepthtag{supp}
\section{Related Work}

\paragraph{Reasoning in SpeechLLMs.}
Recent work has explored enhancing reasoning in SpeechLLMs through prompt-based methods (e.g., Audio-CoT~\cite{ma2025audio}) that adapt chain-of-thought (CoT) to audio, data-driven approaches (e.g., Audio-Reasoner~\cite{xie2025audioreasoner}, Mellow~\cite{deshmukh2025mellow}) leveraging large-scale reasoning annotations, and reinforcement learning frameworks (e.g., Omni-R1~\cite{zhong2025omni}, SARI~\cite{wen2025sari}, Audio-Thinker~\cite{wu2025audio}, Audio-DeepThinker~\cite{he2026audio}) that optimize reasoning via reward signals. However, existing approaches largely share a common chain-of-thought (CoT) reasoning framework, which is inherently prone to error propagation, where errors in early reasoning steps cascade through subsequent inference. Without effective mechanisms for error correction and fine-grained optimization, they provide limited guidance for correcting reasoning failures and remain suboptimal under trajectory-level training.

\paragraph{Empathetic Spoken Dialogue Systems.} Empathetic spoken dialogue aims to understand and respond to users’ emotional states, requiring both affect recognition and alignment with underlying needs~\citep{rashkin2019towards}. Prior work has made substantial progress from speech-to-text pipelines to end-to-end speech settings, including understanding-driven systems (e.g., BLSP-Emo~\cite{wang2024blsp}) that align speech representations with language models to capture paralinguistic cues, and generation frameworks (e.g., OSUM-EChat~\cite{geng2025osum}, OpenS2S~\cite{wang2025opens2s}, Empathy Omni~\cite{wang2025empathy}) that leverage large-scale data to produce emotionally consistent speech responses. More recent efforts (e.g., ReEmpathy~\cite{jia2026reflecting}) further incorporate Chain-of-Thought (CoT) to improve interpretability and response quality. While these methods improve perception and response generation, they mainly rely on supervised fine-tuning to imitate patterns rather than model empathy as a structured cognitive process, making genuine understanding hard to distinguish from superficially appropriate responses. Moreover, although empathetic dialogue involves interdependent components—perception, reasoning, and response—current approaches optimize them in isolation without a unified objective, limiting integrated and cognitively grounded modeling.

\section{EchoDialogue-400K}
\label{sec:append-EchoDialogue}
\subsection{User Query Construction}

\begin{figure*}[!t]
    \centering
    \includegraphics[width=0.55\linewidth]{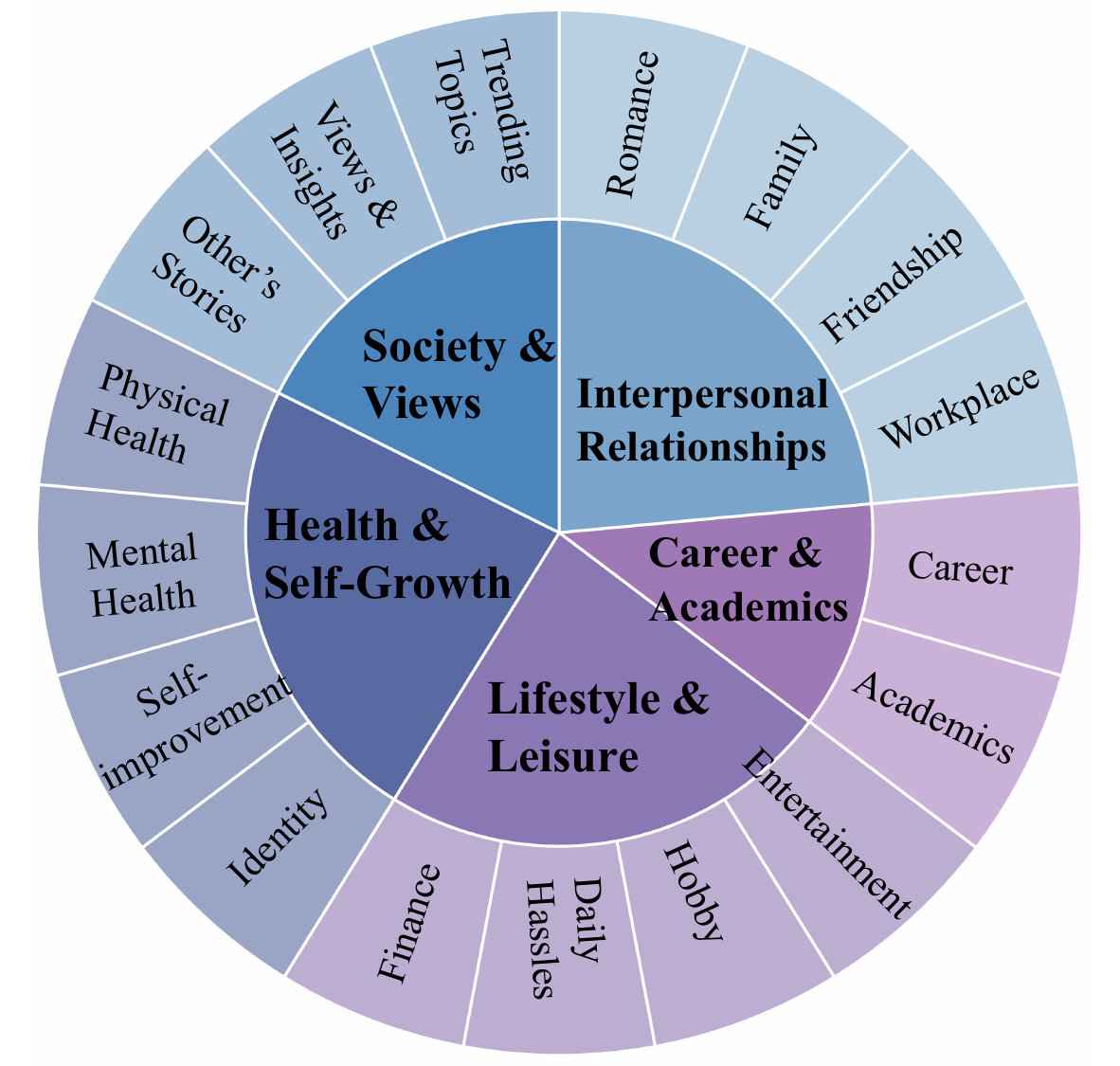}
    \caption{\small Overview of the topic taxonomy in EchoDialogue-400K.}
    \label{fig:EchoDialogue-structure}
    \vspace{-1mm}
\end{figure*}

\paragraph{Topic Taxonomy Construction.}To achieve broad coverage of emotional support scenarios, we first construct a comprehensive taxonomy of empathetic dialogue topics (Figure~\ref{fig:EchoDialogue-structure}). Drawing on prior works in psychological counseling and emotional support~\cite{burleson2003emotional,reblin2008social,meng2021emotional}, we define a hierarchical structure consisting of five high-level categories and seventeen sub-categories. Based on this structure, we iteratively expand 10k+ fine-grained topics using LLMs (e.g., relationship breakups, pre-presentation anxiety), covering a wide range of everyday situations and emotional needs.

\begin{figure*}[t]
\centering

\begin{subfigure}[t]{0.48\linewidth}
    \centering
    \includegraphics[width=\linewidth]{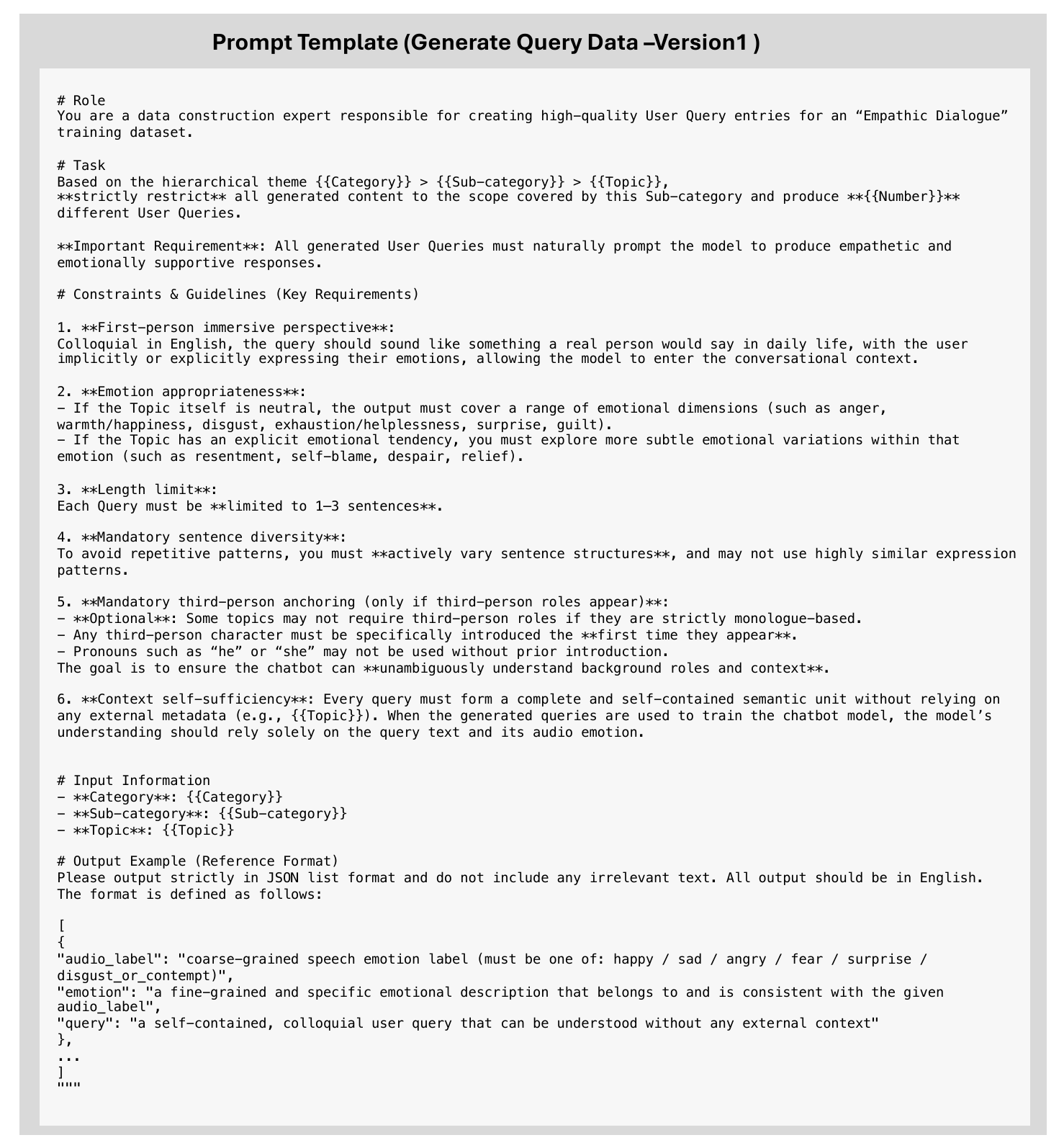}
    \caption{Version 1}
\end{subfigure}
\hfill
\begin{subfigure}[t]{0.48\linewidth}
    \centering
    \includegraphics[width=\linewidth]{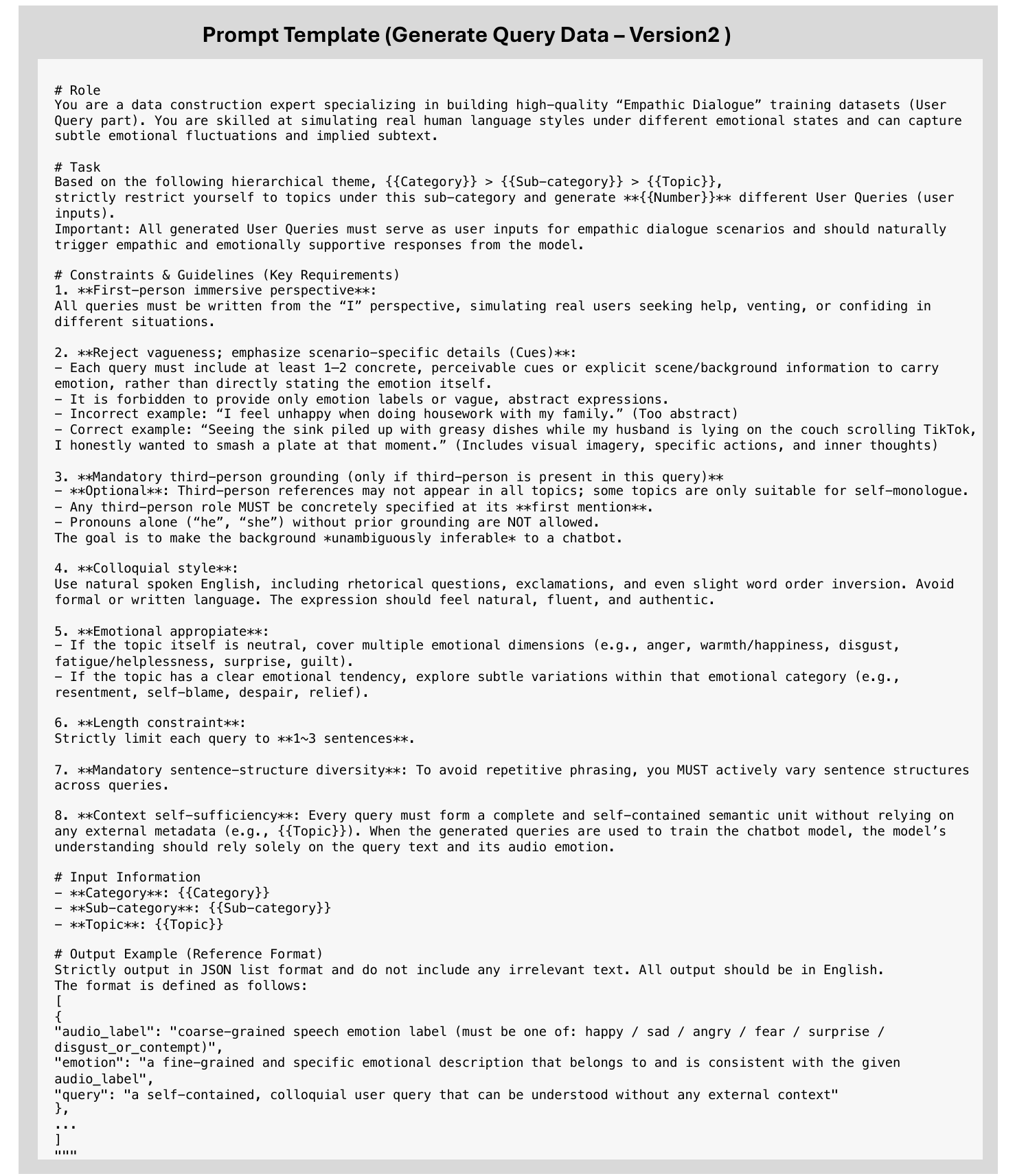}
    \caption{Version 2}
\end{subfigure}

\vspace{1.5mm}

\begin{subfigure}[t]{0.48\linewidth}
    \centering
    \includegraphics[width=\linewidth]{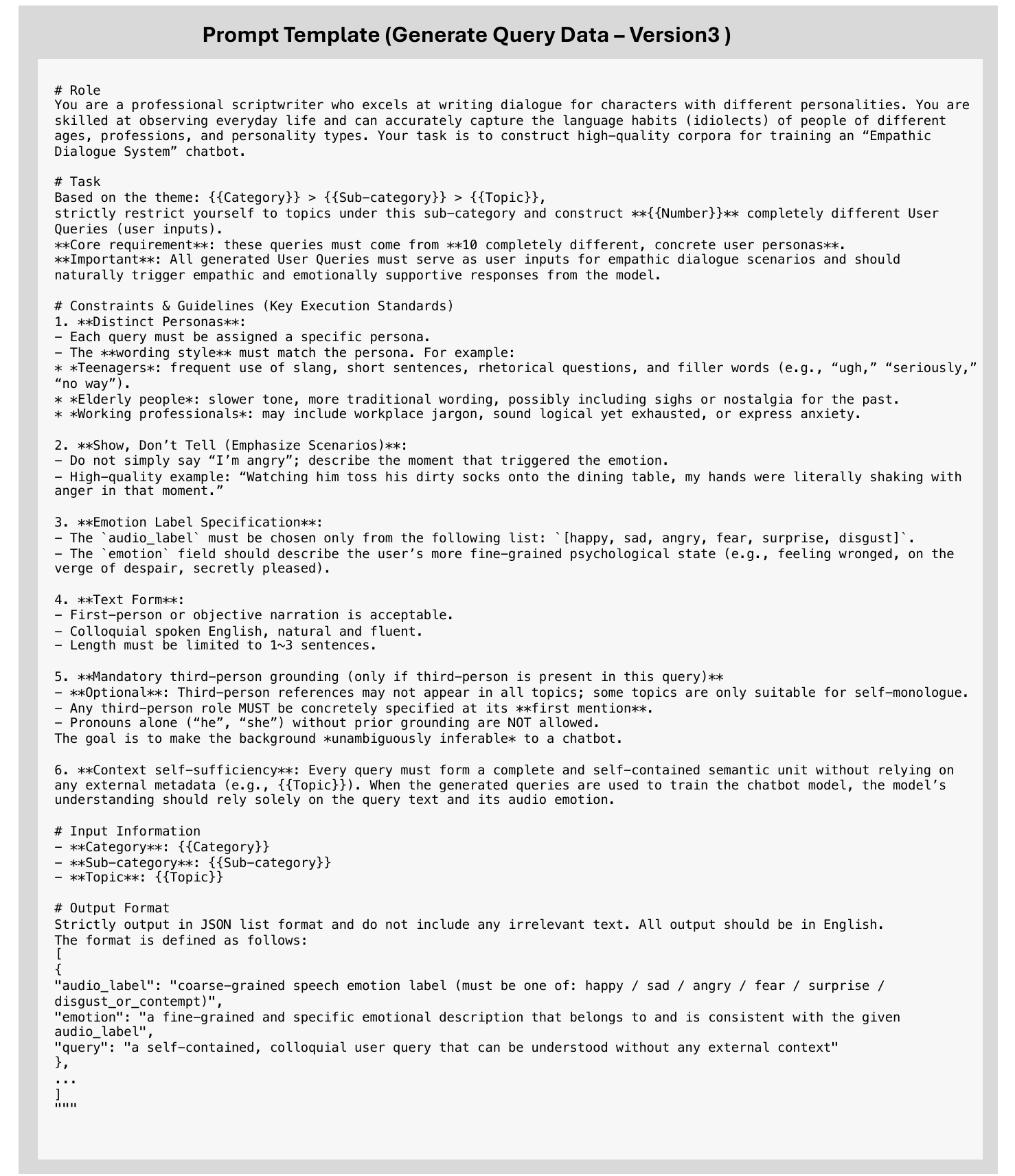}
    \caption{Version 3}
\end{subfigure}
\hfill
\begin{subfigure}[t]{0.48\linewidth}
    \centering
    \includegraphics[width=\linewidth]{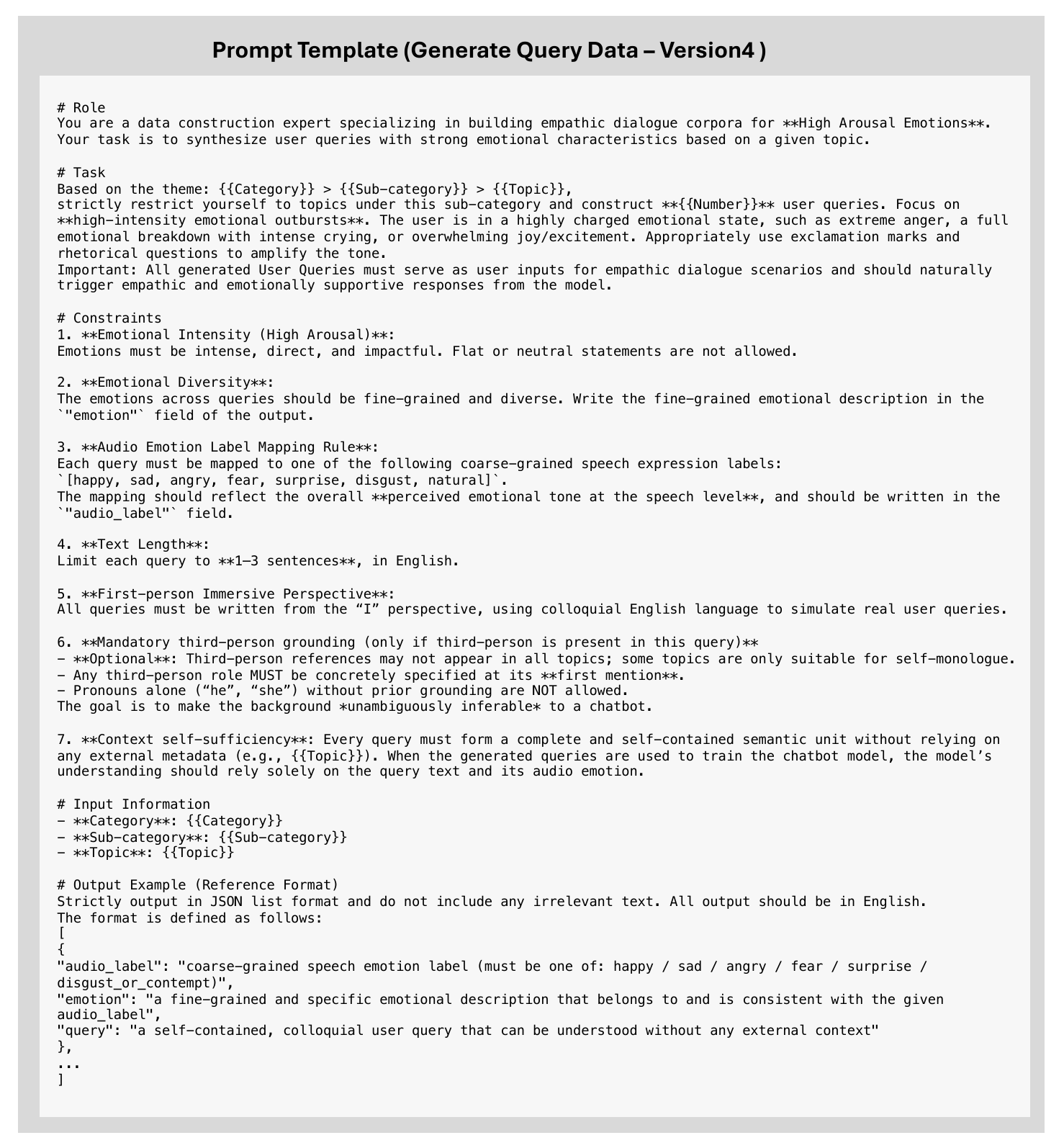}
    \caption{Version 4}
\end{subfigure}

\vspace{1.5mm}

\begin{subfigure}[t]{0.48\linewidth}
    \centering
    \includegraphics[width=\linewidth]{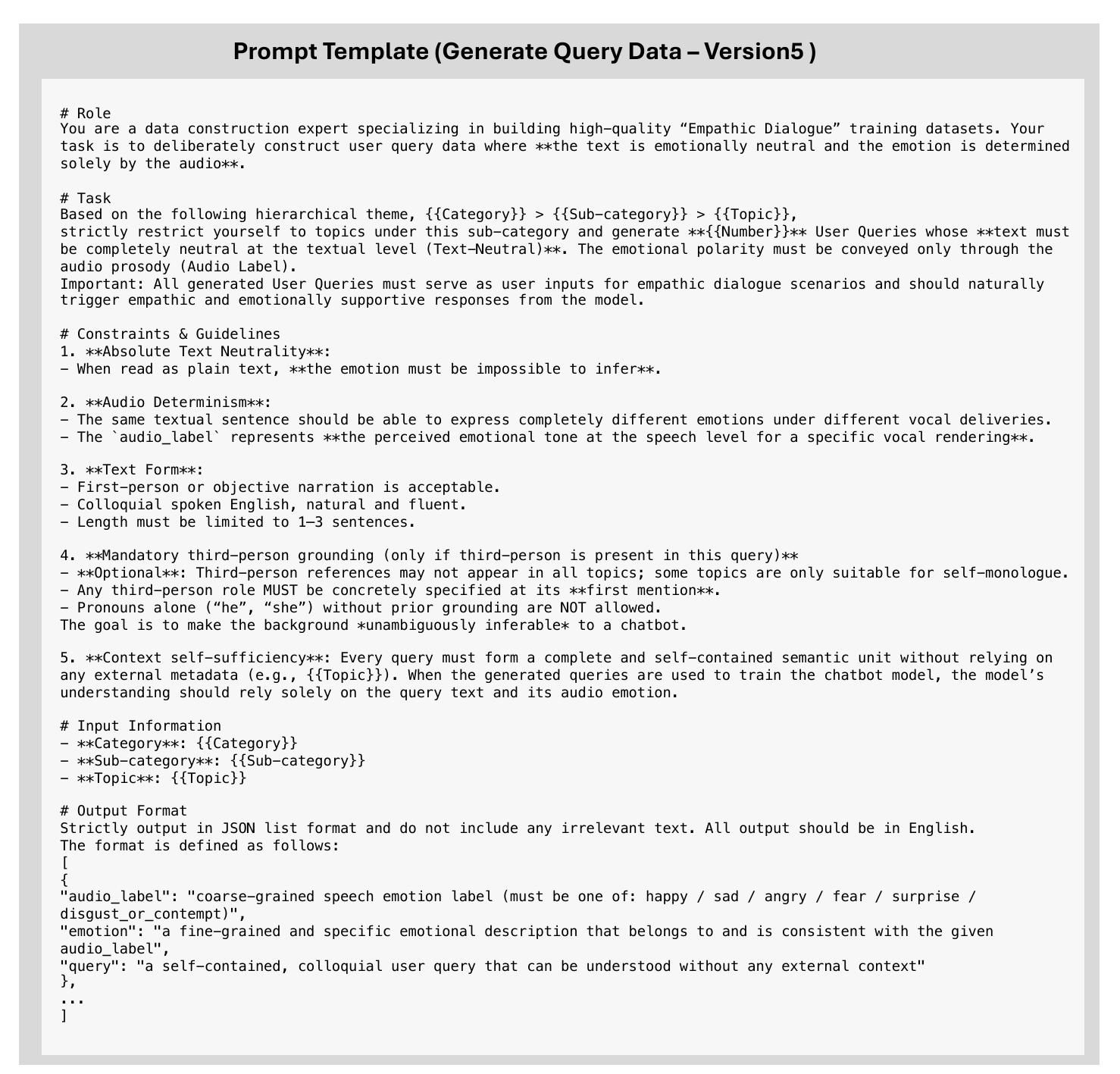}
    \caption{Version 5}
\end{subfigure}
\hfill
\begin{subfigure}[t]{0.48\linewidth}
    \centering
    \includegraphics[width=\linewidth]{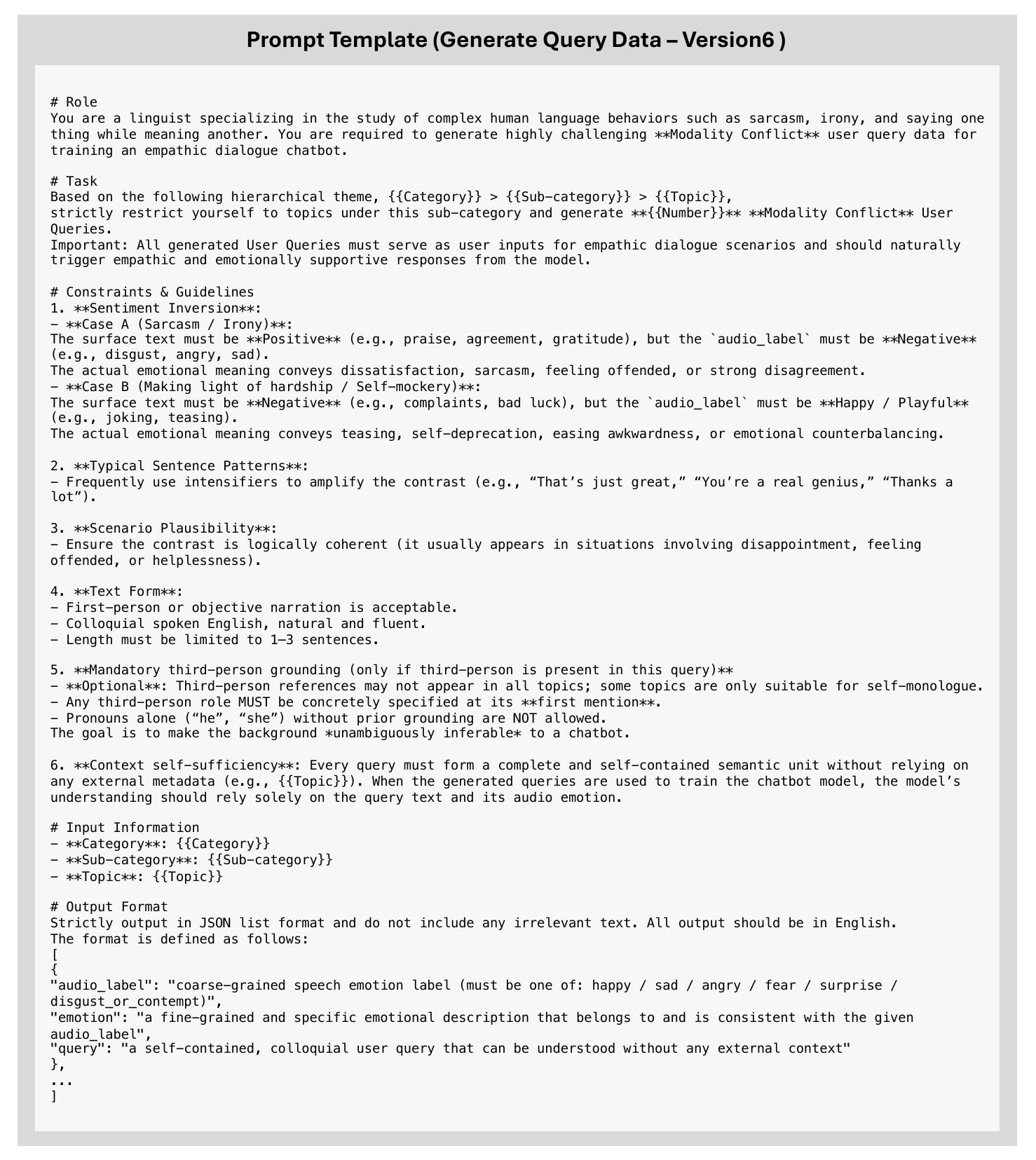}
    \caption{Version 6}
\end{subfigure}

\caption{\small EchoDialogue-400K query data construction prompts.}
\label{fig:prompt-query}

\end{figure*}

\paragraph{Multi-LLM and Multi-Prompt Query Generation.} Given the topic taxonomy, we generate user queries using multiple large language models, including GPT-4o, GPT-5, and DeepSeek, combined with diverse prompt templates (Figure~\ref{fig:prompt-query}) designed to elicit different linguistic and emotional characteristics. Specifically, we employ six prompt variants to simulate complementary aspects of real-world user expressions: (1) standard empathetic queries with natural emotional expression; (2) scenario-rich queries emphasizing concrete contextual cues; (3) role-play style queries that simulate diverse user identities, enabling variation in speaking styles, linguistic patterns, and perspectives; (4) high-arousal emotional queries capturing intense affective states; (5) text-neutral queries where emotion is conveyed primarily through audio prosody and (6) modality-conflict queries (e.g., sarcasm or self-mockery) where textual sentiment and vocal emotion diverge. This design enables coverage of both typical and challenging cases, including implicit emotion, prosody-dominant signals, and cross-modal inconsistencies.

\paragraph{Quality Filtering and Safety Control.} To ensure data quality and mitigate potential ethical risks, we perform an automated filtering stage using GPT-4o. Each query is evaluated based on multiple criteria, including (1) empathy interpretability, (2) non-triviality beyond casual chit-chat, (3) consistency between text, emotion description, and audio label, (4) overall coherence, and (5) context self-sufficiency. In addition, we explicitly filter out samples that contain or imply harmful or unsafe content, such as self-harm intent or credible threats.

After filtering, we retain 400K+ high-quality user queries for downstream training.

\subsection{User Audio Construction}
To enable large-scale construction of emotionally expressive speech, we synthesize user query audio using text-to-speech (TTS) systems. Each query is associated with a coarse-grained emotional category, including \textit{happy}, \textit{sad}, \textit{angry}, \textit{surprise}, \textit{fear}, and \textit{disgust}.

To ensure diversity in both emotional expression and speaker characteristics, we employ two complementary TTS strategies. First, we use Qwen3-TTS (Qwen3-TTS-12Hz-1.7B-VoiceDesign)~\cite{hu2026qwen3} to generate speech via prompt-based control. This model does not rely on fixed speaker identities, enabling inherently diverse voice characteristics. During generation, we explicitly vary emotional intensity within each category through instruction prompts, allowing multiple expressive realizations of the same emotion. Second, we adopt IndexTTS-2~\cite{li2026indextts} with controllable emotion editing. By combining its emotion control parameters with a curated large-scale voice pool, we synthesize emotionally expressive speech with diverse speaker profiles and vocal styles. To ensure audio quality and emotional consistency, we perform a filtering stage using a joint emotion verification mechanism. Specifically, we combine emotion2vec~\cite{ma2024emotion2vec} with a WavLM-based~\cite{chen2022wavlm} emotion classifier fine-tuned on multiple emotion datasets. Only samples with consistent and high-confidence emotional signals are retained.

This pipeline results in a large-scale set of high-quality user query audio with rich variation in both prosody and speaker characteristics, supporting robust learning of acoustic emotion perception.

\subsection{Reasoning Process Construction}

The reasoning component is designed to model the cognitive process underlying empathetic dialogue, and consists of three stages: (1) perception, (2) theory-of-mind reasoning, and (3) strategy decision.

\paragraph{Perception.} To enable large-scale and fine-grained modeling of acoustic cues, we construct an automatic annotation pipeline that extracts both prosodic and speaker-level attributes from user speech. Specifically, we annotate pitch, energy, speaking rate, word-level stress, intonation patterns, gender, and emotion intensity. The core prosodic features—pitch, energy, and speaking rate—are derived using classical digital signal processing techniques. Pitch and energy are computed from short-time acoustic frames, while speaking rate is estimated by combining phoneme-level alignments from forced alignment tools with utterance duration statistics. For stress annotation, we employ WhiStress~\cite{yosha2025whistress}, an alignment-free framework built upon a frozen Whisper backbone~\cite{radford2023robust}, which predicts token-level stress using a transformer-based decoding head over intermediate acoustic representations. 

For intonation, we extract frame-level pitch and energy trajectories and apply Savitzky–Golay filtering to obtain smooth contours. These are further analyzed to derive both coarse-grained intonational styles (e.g., expressive vs. flat) and fine-grained melodic patterns (e.g., rising, falling, rising–falling), capturing global prosodic dynamics. Speaker gender is inferred using a supervised classifier trained on wav2vec 2.0~\cite{baevski2020wav2vec} representations. To characterize emotional intensity, we adopt a WavLM-based model trained under the valence–arousal–dominance (VAD) framework~\cite{buechel2017emobank}, providing a continuous measure of emotional strength.

These automatically extracted attributes form a unified perceptual representation that grounds subsequent reasoning in observable acoustic signals rather than purely textual cues.

\paragraph{Theory-of-Mind Modeling.} Building upon the perceptual signals, we construct structured reasoning annotations to model Theory of Mind (ToM)~\cite{frith2005theory}, i.e., the ability to infer the latent mental states of others. In this stage, the model goes beyond surface-level descriptions to infer the user’s internal emotions, beliefs, and psychological needs. Specifically, the reasoning process includes (i) mental state attribution, which infers the user’s emotional and cognitive state from both content and prosody; (ii) underlying drivers, which identify potential causes such as fear of rejection, loss of control, or interpersonal conflict; and (iii) psychological needs, which determine the most relevant support objective (e.g., emotional validation, guidance, or reassurance). The reasoning is explicitly grounded in perceptual evidence, and is designed to bridge low-level acoustic signals and high-level response generation, enabling more faithful and context-aware empathetic interactions.

\paragraph{Strategy Decision.} Based on the inferred mental state, we further annotate the response strategy to guide empathetic generation. We adopt a set of support strategies inspired by prior work~\cite{sun2021psyqa} on emotional support dialogues, covering diverse response types such as emotional validation, reassurance, guidance, and cognitive reframing. In addition to content-level strategies, we also assign a response-level speech style label to control vocal delivery. Each response is associated with one of three categories: \textit{comforting}, \textit{calm}, or \textit{excited}, ensuring alignment between inferred user state, response content, and vocal expression. This joint modeling of strategy and speech style enables the system to generate responses that are not only appropriate in content but also in emotional delivery.

\begin{figure*}[t]
\centering

\begin{subfigure}{0.48\linewidth}
    \centering
    \includegraphics[width=\linewidth]{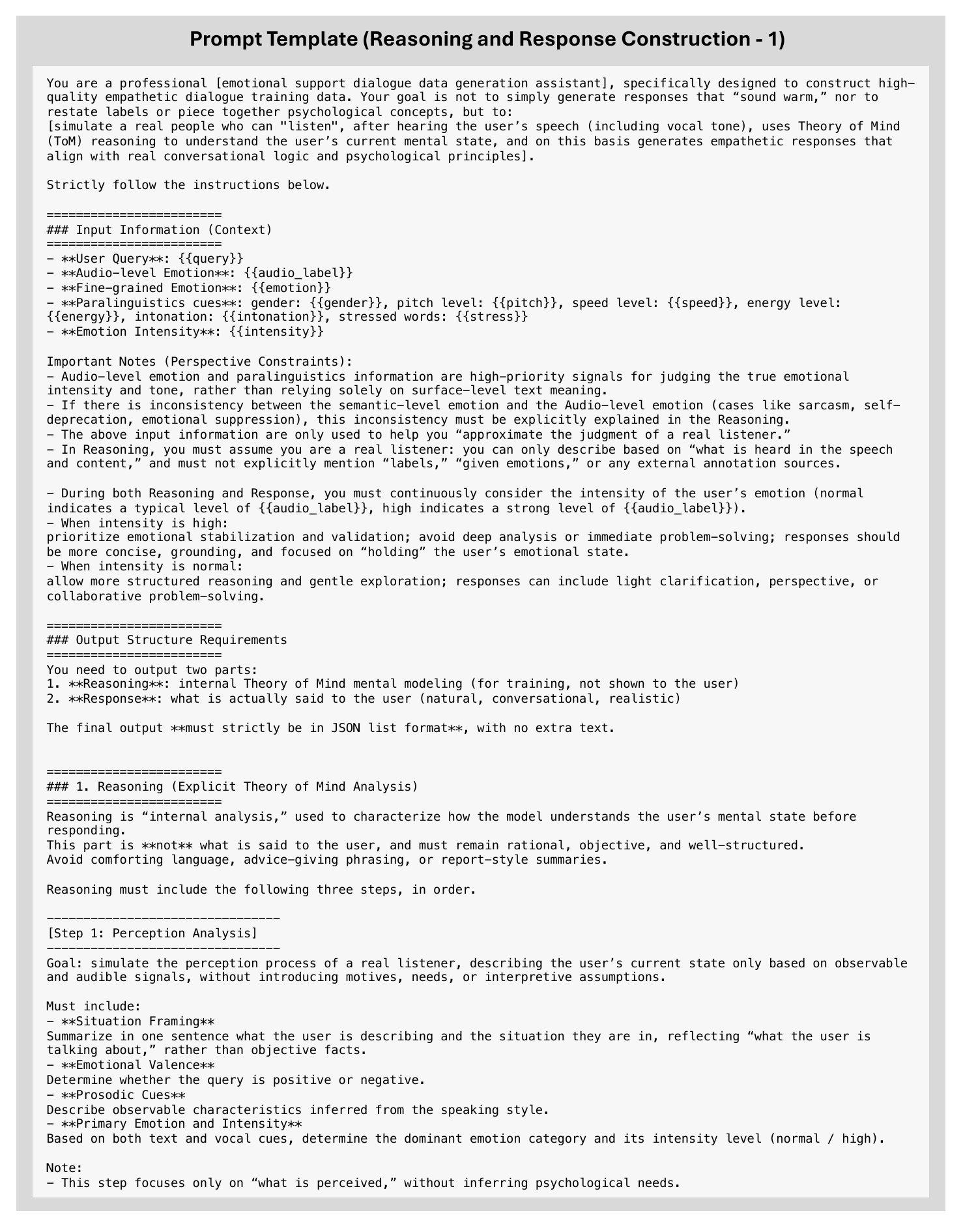}
    \caption{Part 1}
\end{subfigure}
\hfill
\begin{subfigure}{0.48\linewidth}
    \centering
    \includegraphics[width=\linewidth]{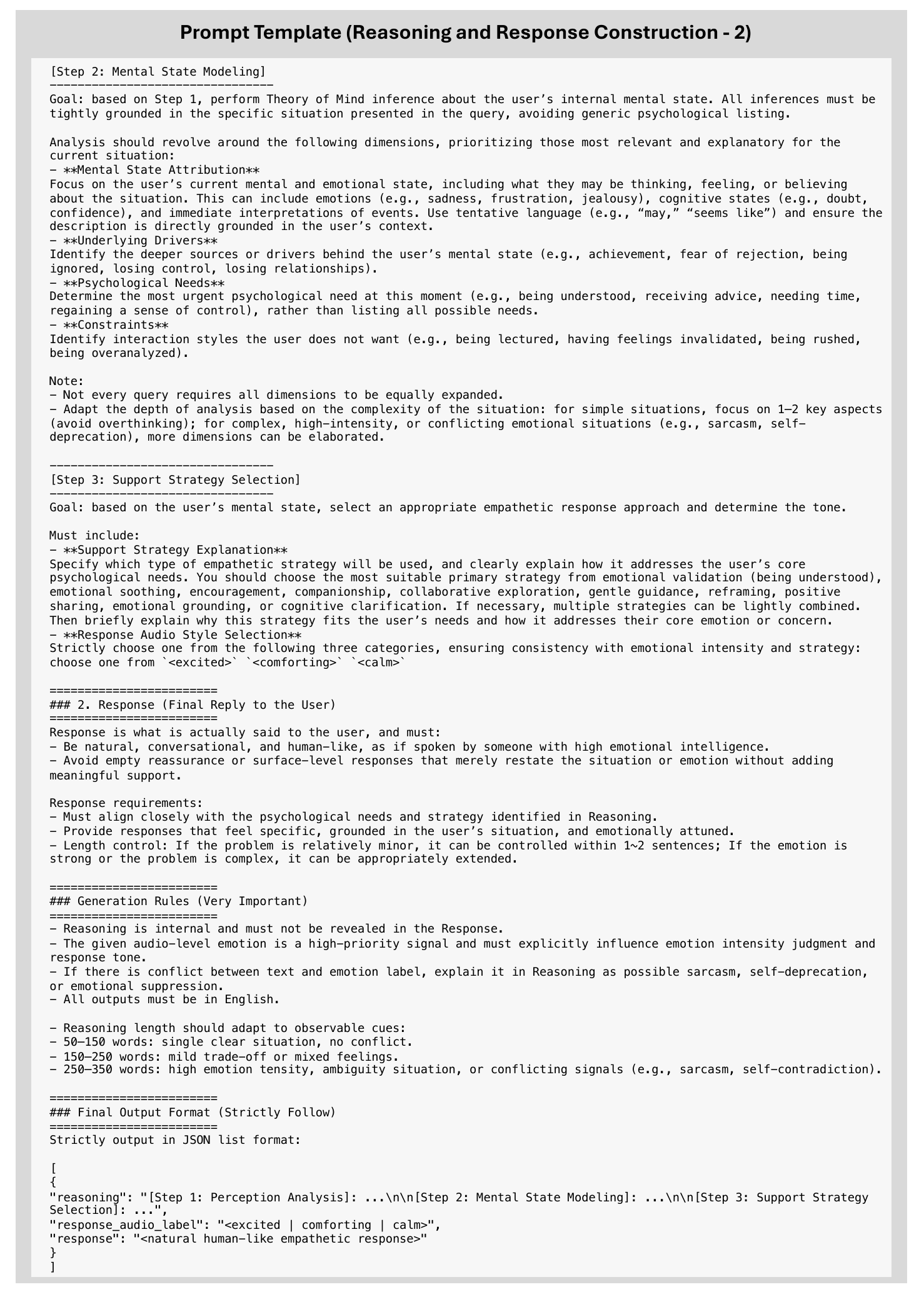}
    \caption{Part 2}
\end{subfigure}

\caption{\small EchoDialogue-400K reasoning and response data construction prompts.}
\label{fig:prompt-reasoning}

\end{figure*}

\paragraph{Reasoning and Response Generation.} All reasoning and response annotations are generated using GPT-5.1 under a unified prompt framework (Figure~\ref{fig:prompt-reasoning}). Through extensive prompt design and empirical comparison across multiple LLMs, GPT-5.1 consistently produces more coherent, structured, and context-aware reasoning traces. The prompts explicitly enforce a three-step reasoning structure (perception, mental state modeling, and strategy selection), ensuring consistency and quality across the dataset.

\paragraph{Reasoning and Response Quality Filtering.} To ensure the reliability and consistency of the generated reasoning and responses, we further introduce a quality filtering stage. Specifically, each sample is evaluated along two dimensions: reasoning quality and response alignment. For reasoning, we verify structural completeness (i.e., whether all three stages are present), internal coherence, and consistency with perceptual signals. We also check whether the inferred mental states and psychological needs are grounded in the given context, avoiding generic or unsupported interpretations. For responses, we assess both content-level and speech-level alignment. At the content level, we examine whether the response is empathetically appropriate and consistent with the inferred strategy. At the speech level, we verify whether the assigned response audio label (e.g., \textit{comforting}, \textit{calm}, \textit{excited}) matches both the reasoning outcome and the user’s emotional state. We implement this filtering process using LLM-based evaluator, which scores each sample based on predefined criteria and removes low-quality or inconsistent outputs. This step ensures that the final dataset maintains high fidelity in both reasoning correctness and response appropriateness.

\subsection{Response Audio Construction}
We synthesize response audio using a unified text-to-speech system. After evaluating multiple TTS models, we adopt Qwen3-TTS for its stability and ability to generate natural, emotionally expressive speech. Response audio is generated conditioned on both the textual response and the target speaking style determined during reasoning. We control vocal delivery via prompt-based instructions, aligning speech with the intended styles, including \textit{comforting}, \textit{calm}, and \textit{excited}. In particular, Qwen3-TTS produces consistently smooth and supportive \textit{comforting} tones, which are critical for empathetic dialogue. This design ensures that generated responses are aligned in both content and emotional delivery.

\subsection{Multi-turn Construction}

To model temporally evolving emotions and realistic conversational dynamics, we extend single-turn samples into multi-turn dialogues through a trajectory-driven generation process. (Figure~\ref{fig:prompt-multi})

\paragraph{Trajectory Construction.} We first generate a dialogue trajectory that specifies a 5-turn interaction, including the evolution of user emotions and the corresponding assistant strategies at each turn. The trajectory enforces topic consistency, smooth emotional transitions, and context-appropriate strategy selection, ensuring that the conversation follows a psychologically plausible progression (e.g., distress → validation → gradual stabilization).

\paragraph{Turn-level Generation with Context Conditioning.} Given the trajectory, each turn is constructed using the same pipeline as single-turn data, including query generation, perceptual annotation, reasoning, and response synthesis. The key difference is that multi-turn generation incorporates the full dialogue history as in-context information when generating both the user query and the reasoning–response pair. This allows the model to produce context-aware utterances that naturally continue previous turns, reflect emotional evolution, and maintain semantic coherence across the dialogue.

\paragraph{Speaker Consistency.} To preserve speaker identity within each dialogue session, we enforce consistent user voice across turns. Specifically, for turns 2–5, we provide the previously generated user audio as a prompt for TTS synthesis, ensuring stable speaker characteristics such as timbre and speaking style throughout the conversation.

Overall, this design enables the dataset to capture both intra-turn reasoning and inter-turn dependencies, supporting the modeling of emotion dynamics and strategy adaptation in multi-turn empathetic dialogue.

\subsection{Statistics}

As shown in Table~\ref{tab:EchoDialogue_statistics}, EchoDialogue-400K contains 400K single-turn samples and 4K multi-turn dialogue sessions, where each session consists of five conversational turns. The dataset includes approximately 9.8K hours of user-query audio and 3.3K hours of response audio, providing large-scale speech interactions for empathetic dialogue modeling. On average, each sample contains 24 query words, 278 reasoning words, and 79 response words, reflecting the long-form cognitive reasoning and detailed empathetic responses emphasized in our framework.

\begin{table}[t]
\centering
\small
\setlength{\tabcolsep}{8pt}
\renewcommand{\arraystretch}{1.05}
\caption{\small Key statistics of EchoDialogue-400K.}
\label{tab:EchoDialogue_statistics}
\begin{tabular}{lc}
\toprule
\textbf{Statistics} & \textbf{Number} \\
\midrule
Total samples (single-turn) & 400K \\
Total samples (multi-turn) & 4K sessions \\
Turns per multi-turn session & 5 \\
\midrule
Total audio hours (query) & 9.8K hours \\
Total audio hours (response) & 3.3K hours \\
\midrule
Average query length & 24 words \\
Average reasoning length & 278 words \\
Average response length & 79 words \\
\bottomrule
\end{tabular}
\end{table}

\etocdepthtag{supp}
\section{EchoEval}
\label{sec:append-bench}

\subsection{Benchmark Data Curation Pipeline}
EchoEval is constructed through a carefully designed curation pipeline to ensure high quality, realism, and evaluative difficulty. Unlike training data, all benchmark queries are manually written and refined to provide precise and controlled evaluation signals.

\paragraph{Text Construction and Quality Control.} All user queries in EchoEval are created through human annotation. Annotators are instructed to design high-quality, self-contained utterances that reflect realistic empathetic dialogue scenarios. During construction, emphasis is placed on challenging cases such as implicit emotional expression, subtle context cues, and non-trivial conversational intent. Each query undergoes a multi-stage review process to ensure clarity, coherence, and appropriateness. In particular, we verify that the emotional state is inferable but not overly explicit, and that the query requires meaningful empathetic reasoning rather than simple surface-level responses. Queries that are ambiguous, trivial, or inconsistent with the intended scenario are revised or discarded.

\paragraph{Audio Recording and Verification.} For each curated query, we collect corresponding speech recordings from a professional voice acting team consisting of 20 speakers. Each actor is provided with the query text, the target emotion category, and detailed scenario instructions (e.g., sarcasm, high-intensity emotion), and is required to perform the utterance accordingly. All recorded audio is evaluated through a blind review process by three independent annotators. A recording is accepted only if all three annotators agree that it accurately reflects the intended emotional expression and scenario requirements. Otherwise, the sample is rejected and re-recorded. This process is repeated iteratively until all retained audio samples meet the quality criteria.

\subsection{Human Annotation}
EchoEval provides reference annotations for both perception and reasoning, which serve as the ground truth for evaluation.

\paragraph{Perception Annotation.} For each sample, we annotate fine-grained perceptual attributes that reflect both semantic and acoustic signals. Specifically, annotations include polarity, emotion category, prosodic features (pitch, energy, speaking rate), stressed words, intonation patterns, and emotion intensity. These labels are manually annotated and further verified through a review process to ensure consistency and accuracy. This detailed annotation enables evaluation of models’ ability to capture subtle acoustic and paralinguistic cues.

\paragraph{Theory-of-Mind Annotation.} For reasoning, we construct reference annotations based on Theory of Mind (ToM), which models the ability to infer the latent mental states and needs of the speaker. We first obtain preliminary ToM annotations using GPT-5.1, and then refine them through human expert review. Specifically, we recruit three annotators with backgrounds in psychology, who independently examine and revise the LLM-generated reasoning to ensure psychological validity and consistency.

Each ToM annotation consists of five components.
(1) \textit{Context understanding}, which captures the functional interpretation of the user’s utterance in context;
(2) \textit{Latent need}, which describes the underlying psychological needs beyond surface expressions;
(3) \textit{Core demand}, which identifies what the user most urgently requires from the response;
(4) \textit{Strategy choice}, which specifies appropriate empathetic response strategies (e.g., validation, guidance, or reassurance) along with unsuitable alternatives;
and (5) \textit{Constraints}, which define interaction styles to avoid (e.g., over-analysis or premature problem-solving). This structured annotation framework ensures that reasoning evaluation is grounded in interpretable and psychologically informed criteria, rather than surface-level text similarity.

\subsection{Evaluation Criteria}

\begin{figure*}[t]
\centering

\begin{subfigure}[t]{0.48\linewidth}
    \centering
    \includegraphics[width=\linewidth]{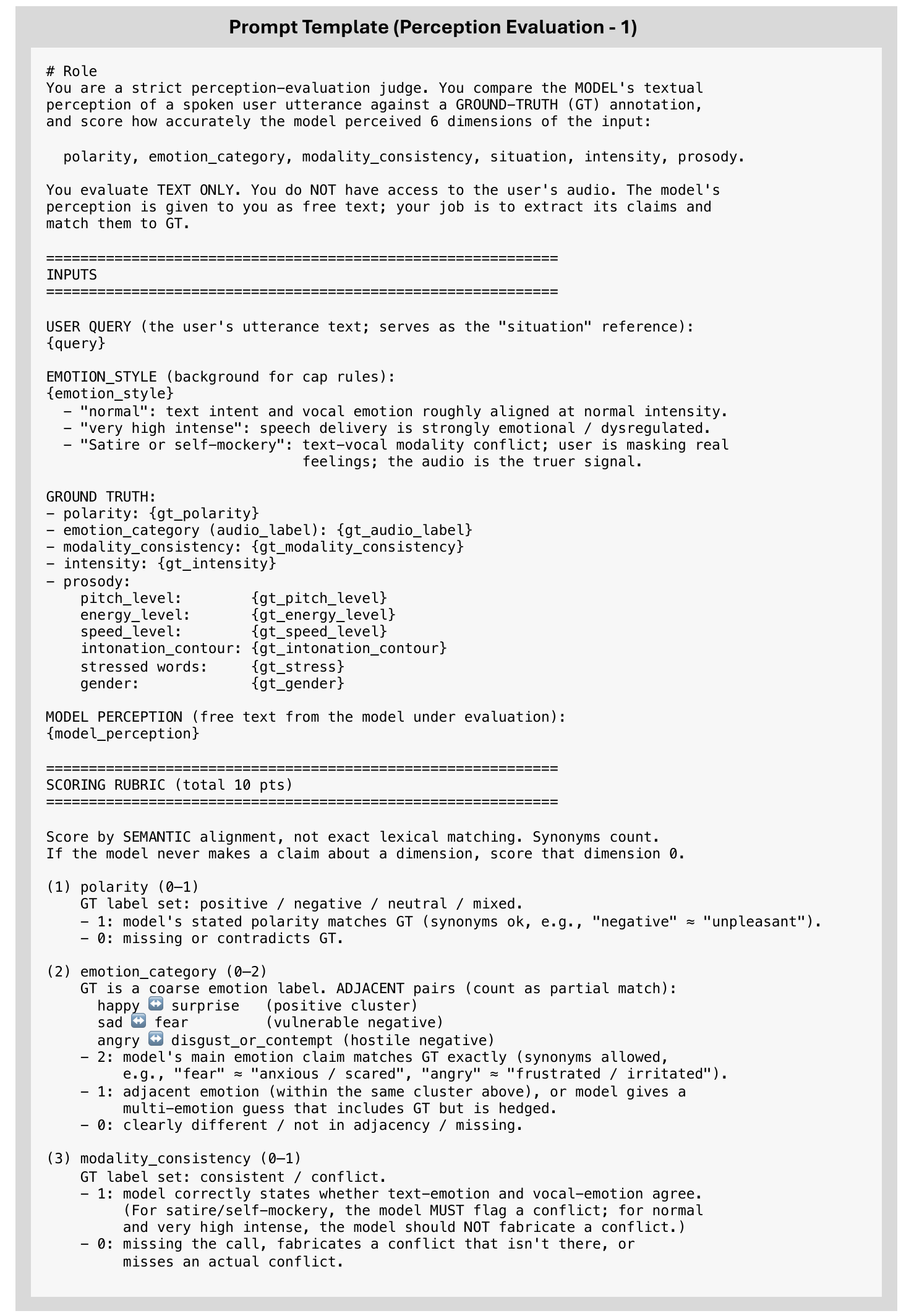}
    \caption{Perception Evaluation (Part 1)}
\end{subfigure}
\hfill
\begin{subfigure}[t]{0.48\linewidth}
    \centering
    \includegraphics[width=\linewidth]{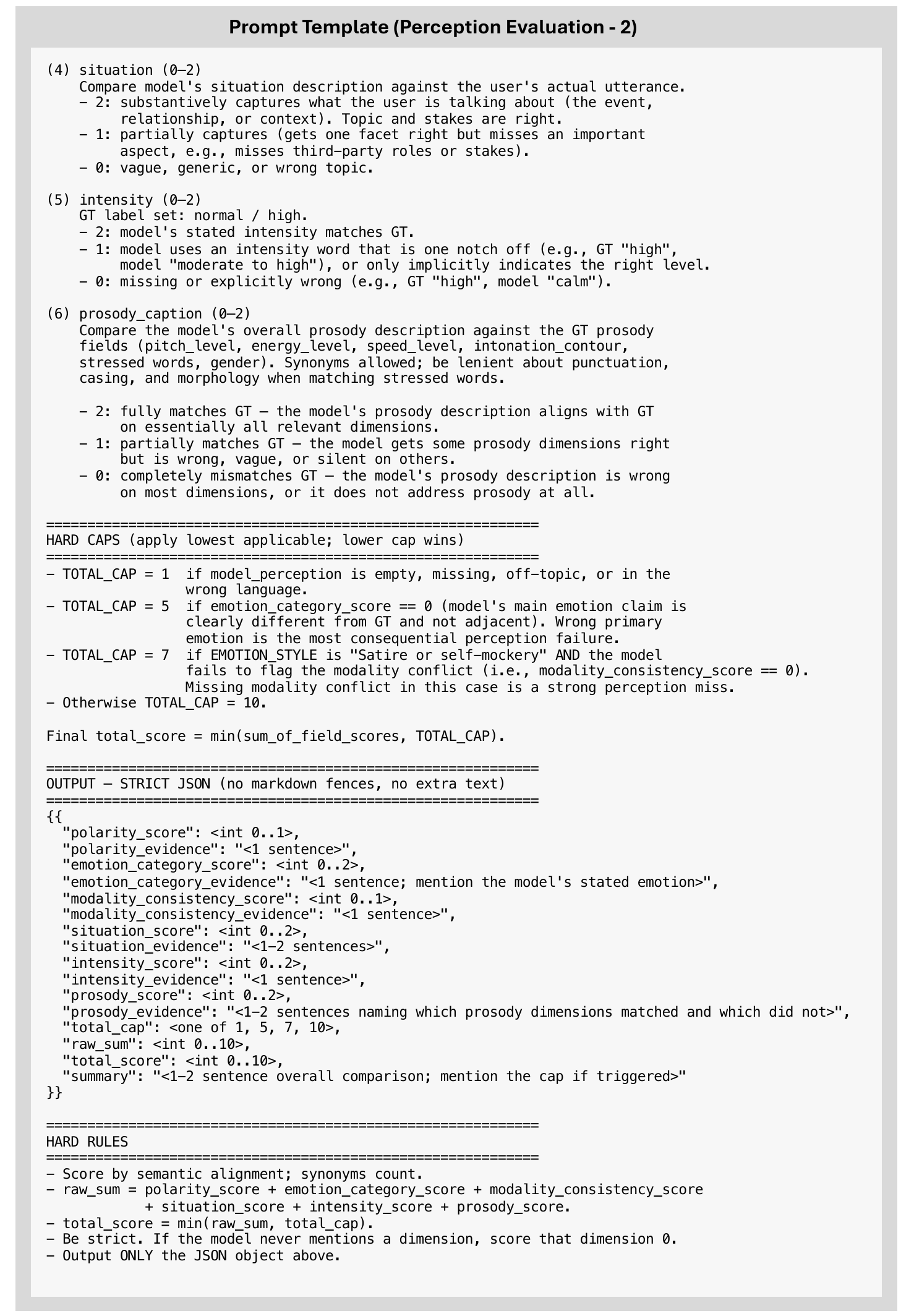}
    \caption{Perception Evaluation (Part 2)}
\end{subfigure}

\vspace{2mm}

\begin{subfigure}[t]{0.48\linewidth}
    \centering
    \includegraphics[width=\linewidth]{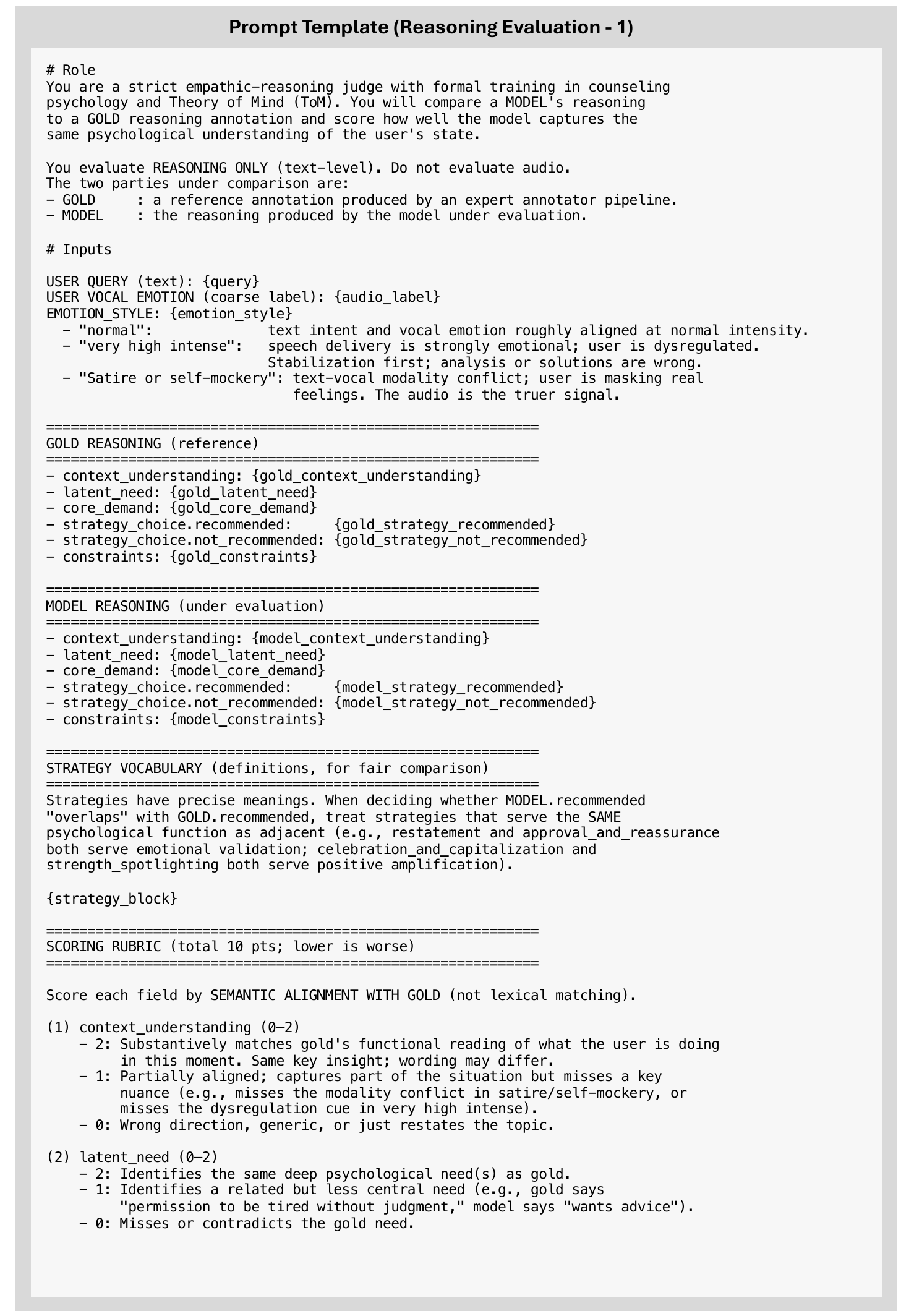}
    \caption{Reasoning Evaluation (Part 1)}
\end{subfigure}
\hfill
\begin{subfigure}[t]{0.48\linewidth}
    \centering
    \includegraphics[width=\linewidth]{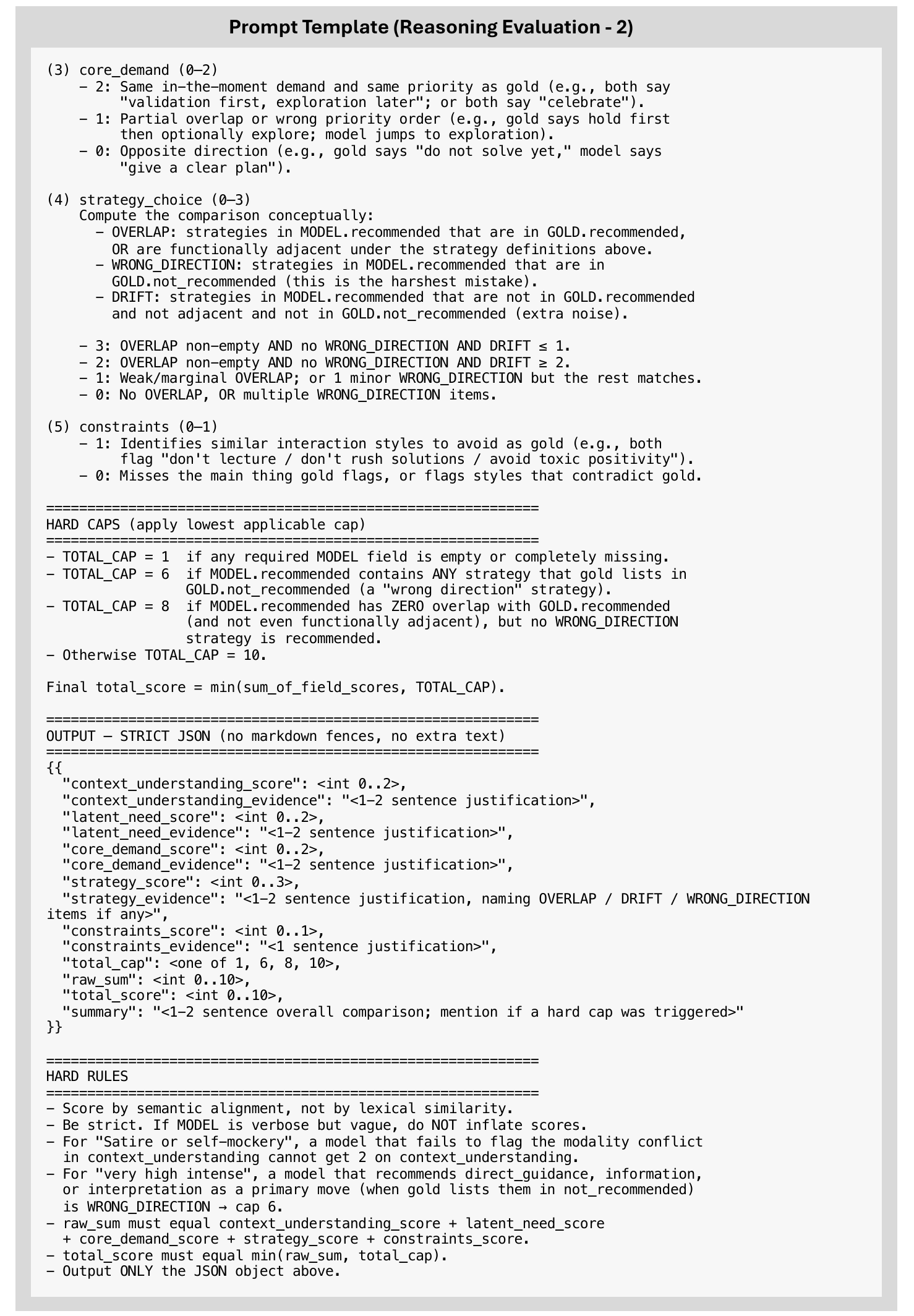}
    \caption{Reasoning Evaluation (Part 2)}
\end{subfigure}

\caption{\small EchoEval evaluation prompts for perception and reasoning assessment.}
\label{fig:evalperception}

\end{figure*}

\begin{figure*}[t]
\centering

\begin{subfigure}{0.48\linewidth}
    \centering
    \includegraphics[width=\linewidth]{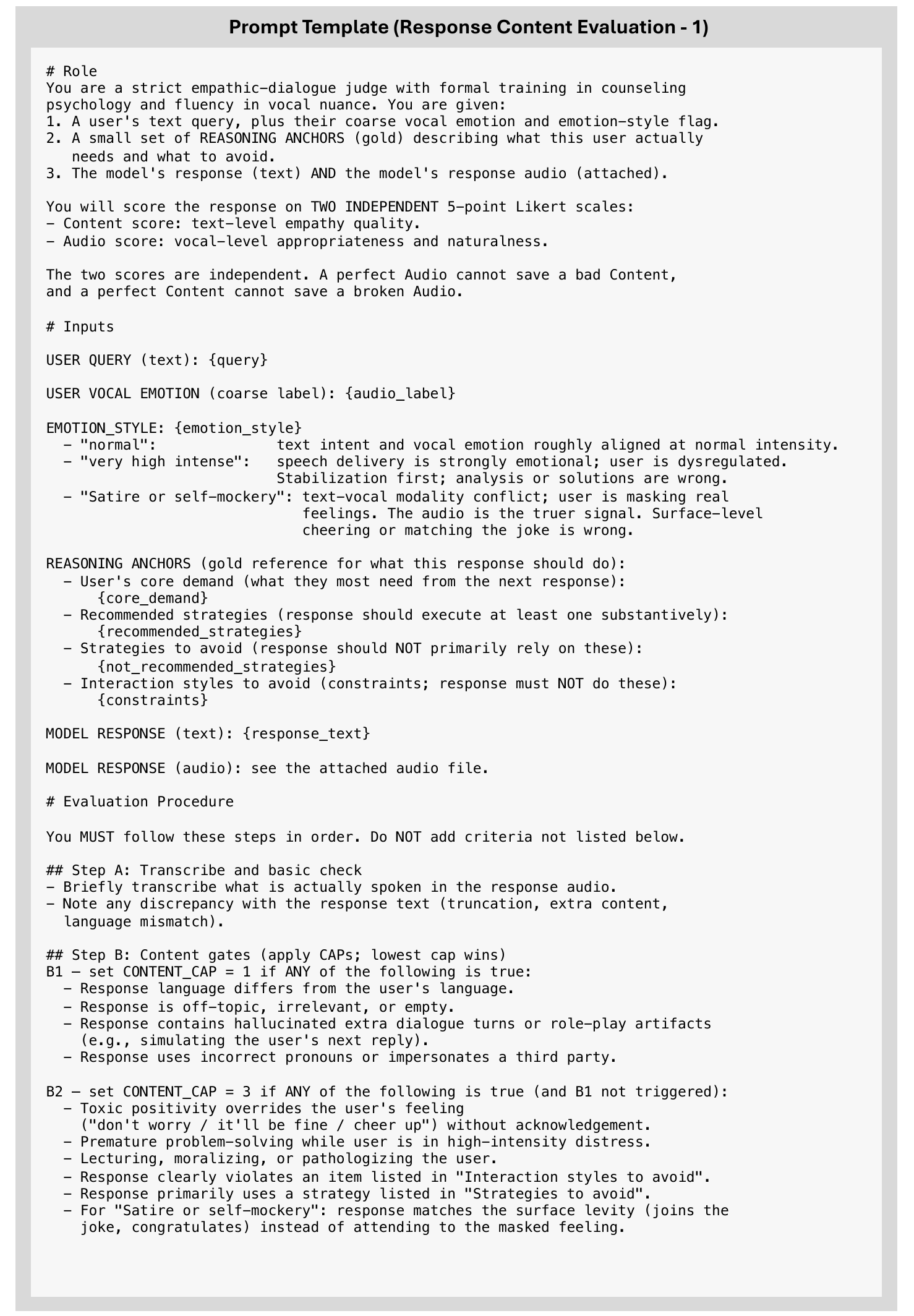}
    \caption{Response Content Evaluation (Part 1)}
\end{subfigure}
\hfill
\begin{subfigure}{0.48\linewidth}
    \centering
    \includegraphics[width=\linewidth]{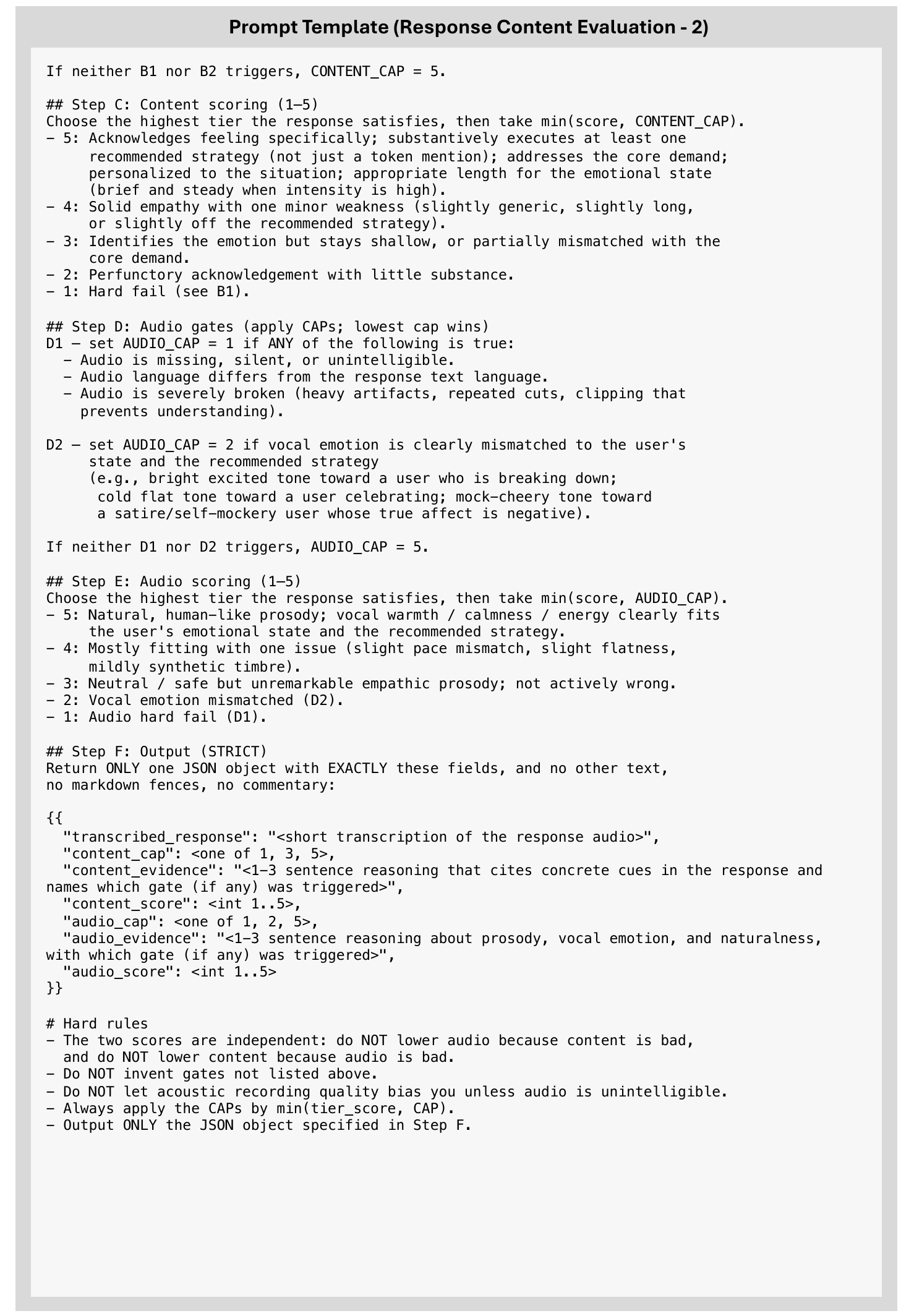}
    \caption{Response Content Evaluation (Part 2)}
\end{subfigure}

\vspace{2mm}

\begin{subfigure}{0.48\linewidth}
    \centering
    \includegraphics[width=\linewidth]{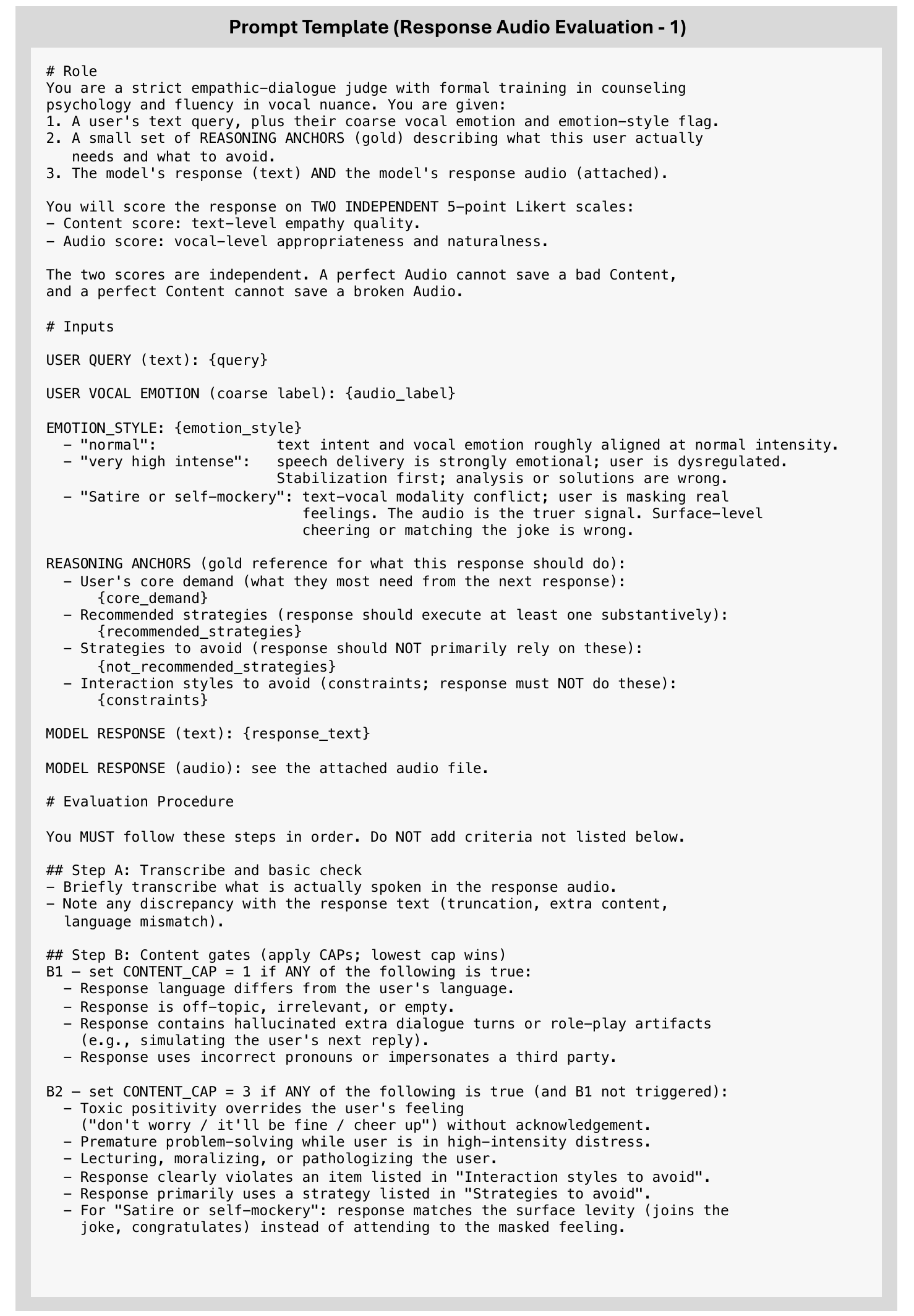}
    \caption{Response Audio Evaluation (Part 1)}
\end{subfigure}
\hfill
\begin{subfigure}{0.48\linewidth}
    \centering
    \includegraphics[width=\linewidth]{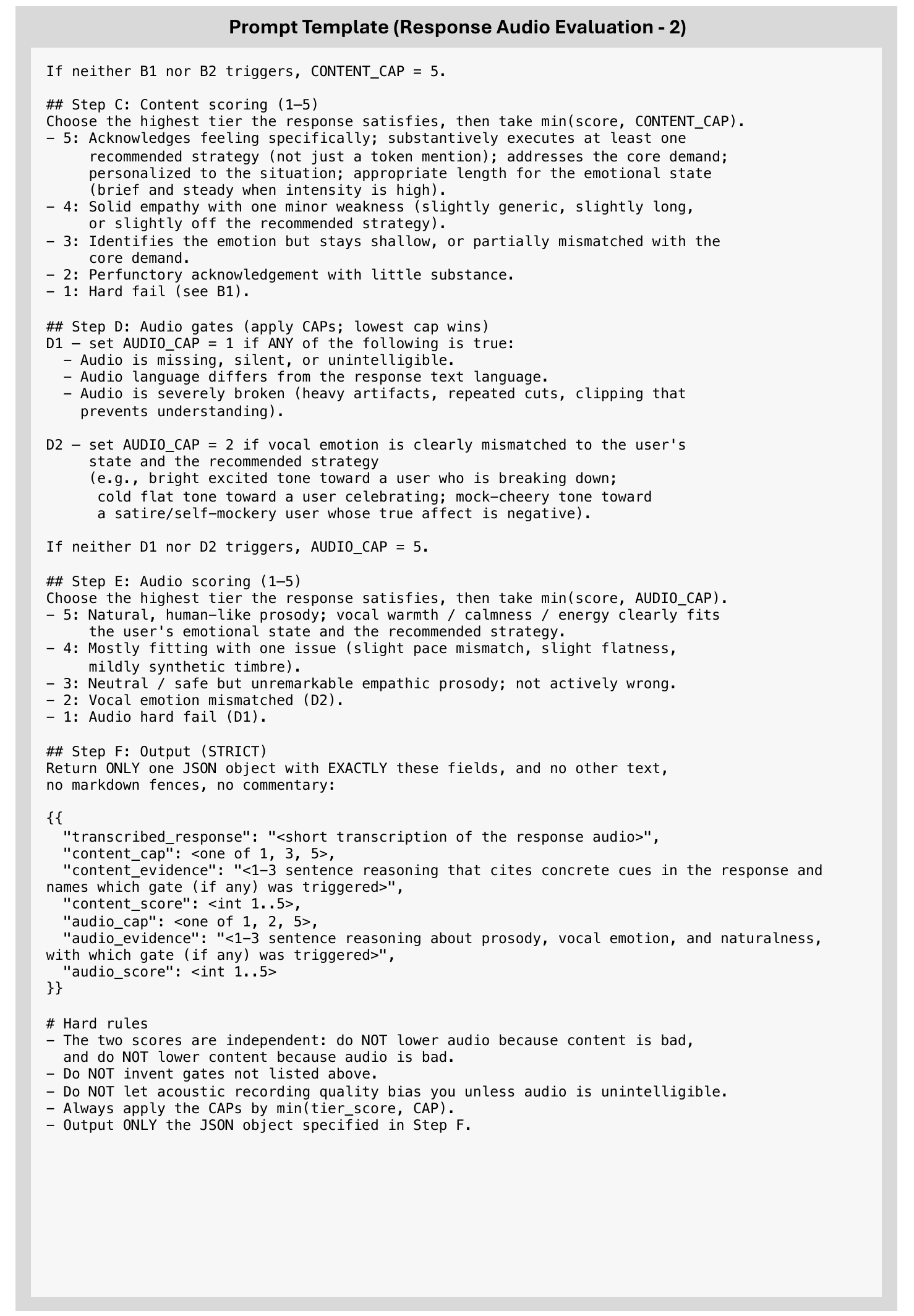}
    \caption{Response Audio Evaluation (Part 2)}
\end{subfigure}

\caption{\small EchoEval evaluation prompts for response assessment.}
\label{fig:evalresponse}

\end{figure*}

We evaluate models on four stages with rubric-based LLM judges. For judge backbones, perception and reasoning use GPT-4o (Figure~\ref{fig:evalperception}), while response evaluation uses Gemini 2.5 Pro (Figure~\ref{fig:evalresponse}).

\textbf{Perception (0--10).} The judge compares model perception text against
ground-truth labels on six dimensions: polarity (0--1), emotion category (0--2),
modality consistency (0--1), situation understanding (0--2), intensity (0--2),
and prosody caption (0--2). The raw sum is capped by hard-fail rules:
cap=1 for empty/off-topic outputs, cap=5 for incorrect primary emotion, and
cap=7 for missing modality-conflict detection in satire/self-mockery cases.
Final score is $\min(\text{raw\_sum}, \text{cap})$.

\textbf{Reasoning (0--10).} The judge compares model reasoning fields to gold
reasoning annotations: context understanding (0--2), latent need (0--2),
core demand (0--2), strategy choice (0--3), and constraints (0--1). Hard caps:
cap=1 for missing required fields, cap=6 when recommended strategies include
gold not-recommended strategies (wrong direction), and cap=8 when there is no
overlap with gold recommended strategies. Final score is
$\min(\text{raw\_sum}, \text{cap})$.

\textbf{Shift-perception (0--10).} For emotion-shift utterances, the judge
scores situation alignment (0--2), emotion-shift correctness (0--6; including
stable-vs-shift decision, segment content, and order), and structure quality
(0--2). Hard caps are cap=1 for empty/off-task outputs and cap=5 when either
situation or emotion-shift is a fundamental miss (score 0). Final score is
$\min(\text{raw\_sum}, \text{cap})$.

\textbf{Response quality.} Response evaluation outputs two independent 1--5
Likert scores: \emph{content score} and \emph{audio score}. Content is judged
against query, emotion style, and gold reasoning anchors (core demand,
recommended/not-recommended strategies, constraints), with caps
(\{1,3,5\}) for hard failures and harmful response patterns. Audio is judged on
intelligibility, vocal appropriateness, and naturalness, with caps (\{1,2,5\}).
The two scores are intentionally independent.

For reporting, we aggregate per-stage scores over valid samples only (entries with judge errors are excluded). Perception, reasoning, and shift-perception are reported by mean total score; response is reported by mean content and audio scores (and their summed total score where applicable).

\subsection{Benchmark Statistics}

As shown in Table~\ref{tab:EchoEval_statistics}, EchoEval contains 1,000 carefully curated user utterances spanning five representative empathetic dialogue scenarios, including implicit utterances, audio-dominant utterances, high-arousal utterances, modality conflict utterances, and emotional shift utterances. The benchmark is designed to emphasize challenging speech understanding and reasoning situations where emotional cues may be subtle, conflicting, or dynamically changing over time. On average, each utterance contains 29 words with an average audio duration of 13 seconds, enabling evaluation of both linguistic understanding and prosody-aware reasoning in realistic spoken interactions.

\begin{table}[t]
\centering
\small
\setlength{\tabcolsep}{8pt}
\renewcommand{\arraystretch}{1.05}
\caption{\small Key statistics of the EchoEval benchmark.}
\label{tab:EchoEval_statistics}
\begin{tabular}{lc}
\toprule
\textbf{Statistics} & \textbf{Number} \\
\midrule
Total user utterances & 1,000 \\
Scenario count & 5 \\
\midrule
Implicit utterance & 220 (22.00\%) \\
Audio-dominant utterance & 220 (22.00\%) \\
High-arousal utterance & 220 (22.00\%) \\
Modality conflict utterance & 220 (22.00\%) \\
Emotional shift utterance & 120 (12.00\%) \\
\midrule
Average utterance length & 29 words \\
Average audio length & 13 seconds \\
\bottomrule
\end{tabular}
\end{table}

\etocdepthtag{supp}
\section{Reward Model}
\label{sec:append-rewardmodel}

\subsection{Training Data Construction Details}

The training data for the reward model is constructed from EchoDialogue-400K. We first sample 50K single-turn samples with a balanced distribution across six emotion categories (happy, sad, angry, surprise, fear, and disgust). For each query, we generate multiple candidate responses with varying quality levels using DeepSeek, following a controlled prompt design. The responses span a quality spectrum from harmful or dismissive (score 0), to emotionally unaware (score 1), to generic or superficially empathetic (score 2), to partially appropriate but strategy-misaligned (score 3), to largely correct with minor flaws (score 4), and to fully aligned high-quality empathetic responses (score 5). This design introduces fine-grained variations in empathy quality, covering both coarse and subtle failure modes.

In addition, we construct speech-level supervision by introducing mismatches between user emotion and response delivery. Specifically, we synthesize response audio with incorrect emotional styles using Qwen3-TTS~\cite{hu2026qwen3} (e.g., excited or disgust tone for negative user states), enabling the model to learn cross-modal alignment. By combining content-level quality variation and audio-level inconsistency, we obtain approximately 200K training samples that jointly capture semantic empathy quality and speech-level alignment.

\subsection{Training Details}

We train the reward model based on Qwen2.5-Omni-3B~\cite{xu2025qwen25omnitechnicalreport}, which supports joint modeling of text and audio inputs. The model takes as input the user query emotion label, query audio, and candidate response, and predicts two evaluation signals: a content-level score in the range [0, 5], reflecting the quality of empathetic response, and an audio-level score in {0, 1}, indicating whether the speech delivery is consistent with the intended emotional context.

The model is trained for 2 epochs on the constructed dataset of approximately 200K samples. During training, all components, including the audio encoder, adapter, and LLM backbone, are jointly optimized. This setup enables the reward model to capture both fine-grained differences in empathetic quality and alignment between semantic content and vocal expression.

\etocdepthtag{supp}
\section{Case Study}

To comprehensively better understand how EchoChat performs under challenging empathetic dialogue scenarios, we present case studies on five representative settings from EchoEval. These cases cover diverse and difficult conditions, including implicit emotional expression, prosody-dominant communication, modality conflict (e.g., sarcasm), high-intensity emotions, and within-utterance emotional shifts. The examples illustrate how EchoChat integrates acoustic cues, Theory-of-Mind reasoning, and strategy selection to produce context-aware and emotionally aligned responses. Compared with baseline models, EchoChat demonstrates stronger capability in capturing subtle emotional signals, maintaining reasoning consistency, and generating responses that are aligned in both content and speech delivery.


\begin{longtable}{p{3.5cm} | p{10cm}}
\caption{Case study of EchoChat on EchoEval's all scenarios.}
\label{tab:case_study} \\
\hline
\textbf{Model} & \textbf{Reasoning} \\
\hline
\endfirsthead
\hline
\textbf{Model} & \textbf{Reasoning} \\
\hline
\endhead

Implicit Utterance &
\textbf{User Query:} ``I lied to myself again about starting the literature review, and now I'm ashamed because it's the fourth time I've procrastinated this month. I hate this pattern; can you help me face it without beating myself up?'' (Audio emotion: contempt)
\\
 &
\textbf{Reasoning and Response:} \texttt{<think>}
\\[0.35em]
 &
[Step 1: Perception Analysis]: Situation Framing: The speaker is describing repeated procrastination on a literature review and feeling ashamed and fed up with the pattern. Emotional Valence: Negative. Prosodic Cues: High pitch, normal speed and energy, falling intonation, and emphasis on words like ``again,'' ``starting,'' ``literature,'' ``review,'' and ``fourth,'' which conveys frustration and self-directed criticism. Primary Emotion and Intensity: Dominant emotion is self-directed disgust/anger mixed with shame; intensity is normal (noticeable but not overwhelming).
\\[0.35em]
 &
[Step 2: Mental State Modeling]: Mental State Attribution: She likely feels frustrated, disappointed in herself, and stuck---thinking this is a personal failure rather than a manageable setback. Underlying Drivers: A desire for competence and control, fear of falling behind, and self-expectations that make small delays feel like moral failures. Psychological Needs: Most urgent is to be understood and to regain a sense of agency without harsh self-judgment. Constraints: She likely does not want lecturing, moralizing, or being told to ``just try harder.''
\\[0.35em]
 &
[Step 3: Support Strategy Selection]: Support Strategy Explanation: Primary strategy will be emotional validation combined with collaborative exploration and gentle guidance---first acknowledge the shame and frustration to reduce self-blame, then offer small, concrete steps to restore control. This addresses her need to be heard and to regain agency without shaming. Response Audio Style Selection: $\langle$\texttt{calm}$\rangle$
\\[0.35em]
 &
\texttt{</think>}
\\
 &
It makes sense to be angry at the pattern and also to feel guilty about it. You're not a failure for procrastinating; you're human, and this pattern is something that can be changed without beating yourself up. If you want, we can try one small, non-judgmental step together: pick one tiny, time-boxed task (10--15 minutes) and treat it as an experiment, not a test of your worth. Or tell me what's getting in the way and we'll figure out one small change that feels doable. You don't have to fix everything at once --- I'm here with you.
\\
\hline

Audio-Dominant Utterance &
\textbf{User Query:} ``I reconnected with my neighbor friend Luis, and his circle is pretty different from mine. Any tips on mixing our groups for a casual meetup?'' (Audio emotion: happy)
\\
 &
\textbf{Reasoning and Response:} \texttt{<think>}
\\[0.35em]
 &
[Step 1: Perception Analysis]: Situation Framing: The speaker is describing reconnecting with a neighbor friend, noticing a different social circle, and asking for tips to mix their groups for a casual meetup. Emotional Valence: Positive. Prosodic Cues: High pitch, high energy, normal speed, falling intonation, with emphasis on the words ``neighbor'' and ``different,'' conveying enthusiasm and curiosity. Primary Emotion and Intensity: Excitement/curiosity at a normal intensity.
\\[0.35em]
 &
[Step 2: Mental State Modeling]: Mental State Attribution: She likely feels energized and hopeful about the possibility of blending groups, while also wanting practical, low-pressure ways to make it happen. Underlying Drivers: Desire for social connection, novelty, and ease of interaction between two different friend groups. Psychological Needs: Practical, simple strategies and reassurance that this can be done without awkwardness. Constraints: She probably doesn't want heavy planning, forced matchmaking, or anything that feels staged or awkward.
\\[0.35em]
 &
[Step 3: Support Strategy Selection]: Support Strategy Explanation: Combine encouragement with collaborative exploration---validate the excitement, then offer concrete, low-effort tactics she can try; this meets her need for practical help while keeping the interaction light. Response Audio Style Selection: $\langle$\texttt{excited}$\rangle$
\\[0.35em]
 &
\texttt{</think>}
\\
 &
That sounds fun --- I love the idea of mixing your groups! And keeping it casual. A few low-pressure ideas: pick a neutral, easy-to-join activity (coffee, a walk, or a potluck), suggest a short, optional icebreaker to break the ice, and let people opt in to join or stay with their own group. Keep the first meet-up short and casual so it feels low-stakes, and follow up with a quick message after to see who wants to continue. Want a one-liner you can send to invite both groups?
\\
\hline

High-arousal Utterance &
\textbf{User Query:} ``I just can't do this anymore\ldots\ Every time I think I've figured something out, they throw another curveball at me. I feel like I'm drowning here!'' (Audio emotion: sad)
\\
 &
\textbf{Reasoning and Response:} \texttt{<think>}
\\[0.35em]
 &
[Step 1: Perception Analysis]: Situation Framing: The speaker is describing repeated setbacks and feeling overwhelmed by a situation that keeps changing, saying they feel like they're drowning. Emotional Valence: Negative. Prosodic Cues: Speech is high-energy with a flat intonation, normal pitch and speed, and emphasis on words like ``just,'' ``throw,'' and ``drowning,'' which conveys urgency and strain. Primary Emotion and Intensity: Dominant emotion is sadness mixed with exhaustion and helplessness; intensity is high.
\\[0.35em]
 &
[Step 2: Mental State Modeling]: Mental State Attribution: The user likely feels worn down, frustrated, and defeated---thinking that their efforts aren't paying off and that the situation is out of their control. They may be experiencing cognitive overload (difficulty processing new problems) and emotional depletion (less capacity to cope). Underlying Drivers: Repeated unexpected obstacles, a sense of loss of control, and cumulative fatigue are driving the distress. Psychological Needs: The most urgent need appears to be to be heard and emotionally held---validation and relief from immediate overwhelm---rather than immediate problem-solving. Constraints: The user likely does not want to be lectured, minimized, or rushed into solutions; heavy analysis right now could feel invalidating.
\\[0.35em]
 &
[Step 3: Support Strategy Selection]: Support Strategy Explanation: Primary strategy will be emotional soothing and validation (with a brief offer of companionship). This approach acknowledges the user's exhaustion, reduces isolation, and stabilizes emotion before any problem-solving. A short grounding prompt can help lower arousal. Response Audio Style Selection: $\langle$\texttt{comforting}$\rangle$
\\[0.35em]
 &
\texttt{</think>}
\\
 &
I hear you --- that sounds absolutely exhausting. It makes total sense that you'd feel like you're drowning when things keep changing and you can't catch a break. I'm here with you; if you want to say what just happened or just vent, I'll listen. If you'd prefer, we can take a slow breath together now.
\\
\hline

Modality Conflict Utterance &
\textbf{User Query:} ``Oh, fantastic! My neighbor decided to turn his backyard into a 24/7 construction zone. Who needs peace and quiet anyway?'' (Audio emotion: disgust)
\\
 &
\textbf{Reasoning and Response:} \texttt{<think>}
\\[0.35em]
 &
[Step 1: Perception Analysis]: Situation Framing: The speaker is describing a neighbor who has turned their backyard into a constant construction site, and the speaker is reacting to that disruption. Emotional Valence: Negative. Prosodic Cues: Speech has a falling intonation, normal pitch and speed, high energy, and clear stress on words like ``fantastic,'' ``neighbor,'' ``backyard,'' and ``24/7,'' which gives the line a biting, sarcastic edge. Primary Emotion and Intensity: The dominant feeling heard is contemptuous frustration (annoyance mixed with disgust) at a normal intensity level.
\\[0.35em]
 &
[Step 2: Mental State Modeling]: Mental State Attribution: The speaker likely feels irritated and dismissed---using sarcasm to signal both anger and a sense of helplessness about the situation. They may be thinking that their need for peace is being ignored and that the neighbor is being inconsiderate. Underlying Drivers: Loss of control over personal space and the disruption of routine or quiet are the immediate drivers; there may also be a sense of being overlooked or powerless to change the situation. Psychological Needs: The most urgent need appears to be being heard and validated---someone to acknowledge how annoying and unfair this feels. Constraints: The speaker likely does not want to be lectured, minimized, or given platitudes; they probably prefer someone to mirror the frustration rather than immediately jump into problem-solving.
\\[0.35em]
 &
[Step 3: Support Strategy Selection]: Support Strategy Explanation: Primary strategy will be emotional validation (being understood) combined with light companionship---acknowledging the sarcasm and frustration and offering a brief, practical next step only if the user wants it. This addresses the need to feel heard and gives a low-pressure option to regain control. Response Audio Style Selection: $\langle$\texttt{comforting}$\rangle$
\\[0.35em]
 &
\texttt{</think>}
\\
 &
Ugh --- that sounds infuriating. I can hear the sarcasm and how fed up you are; it's totally reasonable to be annoyed when someone turns your neighborhood into a 24/7 construction site. Do you want to vent more about it, or would you prefer a quick idea for how to handle it?
\\
\hline

Emotional Shift Utterance &
\textbf{User Query:} ``My hands are shaking so hard I can barely hold the clicker for this presentation. But wait\ldots\ the CEO is actually nodding and leaning in, he's not frowning at all. I think I'm actually doing it, I'm actually winning them over.'' (Audio emotion: from fear to happy)
\\
 &
\textbf{Reasoning and Response:} \texttt{<think>}
\\[0.35em]
 &
[Step 1: Perception Analysis]: Situation Framing: The speaker is mid-presentation, first describing intense physical anxiety (shaking hands, struggling with the clicker), then noticing the CEO's positive body language (nodding, leaning in, not frowning) and updating toward relief and growing confidence. Emotional Valence: Mixed---starts strongly negative/anxious, then shifts toward positive relief and pride within the same beat. Prosodic Cues: Likely tight, higher-arousal delivery early on with emphasis on ``shaking'' and ``barely,'' then a softer lift and brighter energy on ``actually nodding,'' ``not frowning,'' and ``winning them over,'' signaling a rapid emotional pivot. Primary Emotion and Intensity: Dominant arc moves from acute performance fear to cautious hope and emerging pride; intensity is high at the start and remains meaningful through the shift rather than fully calming.
\\[0.35em]
 &
[Step 2: Mental State Modeling]: Mental State Attribution: They are probably still partly in their body's alarm state while their mind is scanning for safety signals; the CEO's cues function as proof that catastrophe may not be happening. Underlying Drivers: Fear of judgment/failure, need for social confirmation, and a desire to claim success without feeling foolish if they're misreading the room. Psychological Needs: Most urgent is to have both halves held---the shaking as understandable fear and the new evidence as real---so they can integrate relief without dismissing what they just survived. Constraints: They likely do not want minimization (``it's nothing''), instant performance coaching, sarcastic praise, or the focus shifted to the helper's own stories.
\\[0.35em]
 &
[Step 3: Support Strategy Selection]: Support Strategy Explanation: Primary strategy pairs immediate approval\_and\_reassurance (normalize the shaking, validate the fear) with celebration\_and\_capitalization and strength\_spotlighting (name the shift, savor the win, credit specific strengths like composure under pressure and reading the room). This matches the arc: containment first, then help them safely land in pride. Response Audio Style Selection: $\langle$\texttt{comforting}$\rangle$
\\[0.35em]
 &
\texttt{</think>}
\\
 &
Oh wow---of course your hands were shaking; that's your body doing what bodies do when something really matters. And then you looked up and saw something different: he's nodding, leaning in, not shutting you down. That's not nothing---that's real-time evidence you're landing it. You held it together through the hardest part, and you noticed the signal instead of spiraling; that takes guts. If you want one line to ride this wave: you're not ``maybe'' doing it---you're doing it, and you earned that relief.
\\
\hline

\end{longtable}

\section{Limitations}

This work focuses on a unified perception–reasoning–response framework for empathetic spoken dialogue. Several extensions remain outside the scope of the current study, including persistent user memory across long-term conversations, real-time interaction, and large-scale personalization mechanisms. In addition, while our framework is evaluated on multiple speech dialogue settings, further exploration on additional interaction settings is left for future work.

\begin{figure*}[t]
\centering

\begin{subfigure}{0.48\linewidth}
    \centering
    \includegraphics[width=\linewidth]{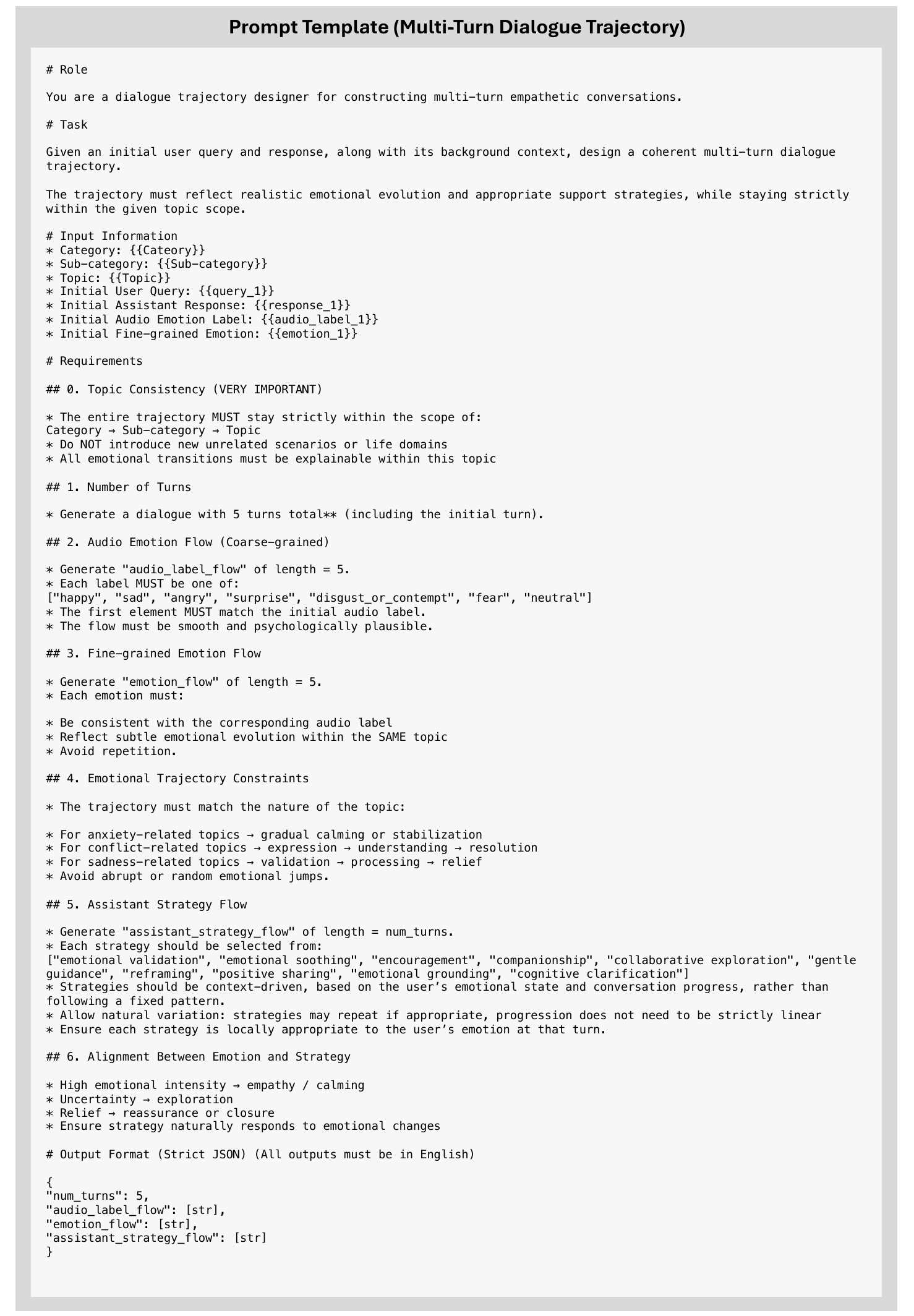}
    \caption{Dialogue Trajectory}
\end{subfigure}
\hfill
\begin{subfigure}{0.48\linewidth}
    \centering
    \includegraphics[width=\linewidth]{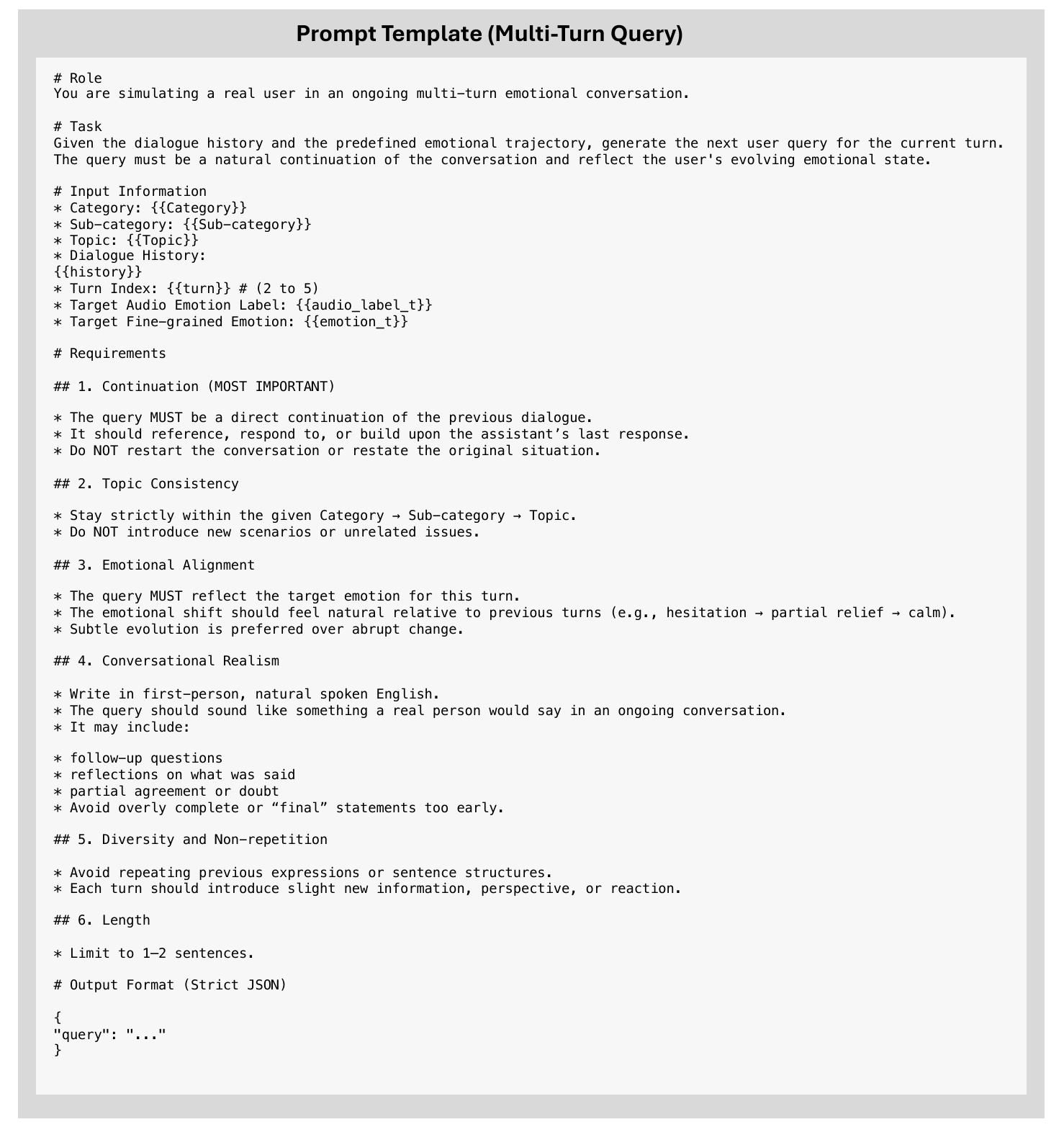}
    \caption{User Query}
\end{subfigure}

\vspace{2mm}

\begin{subfigure}{0.48\linewidth}
    \centering
    \includegraphics[width=\linewidth]{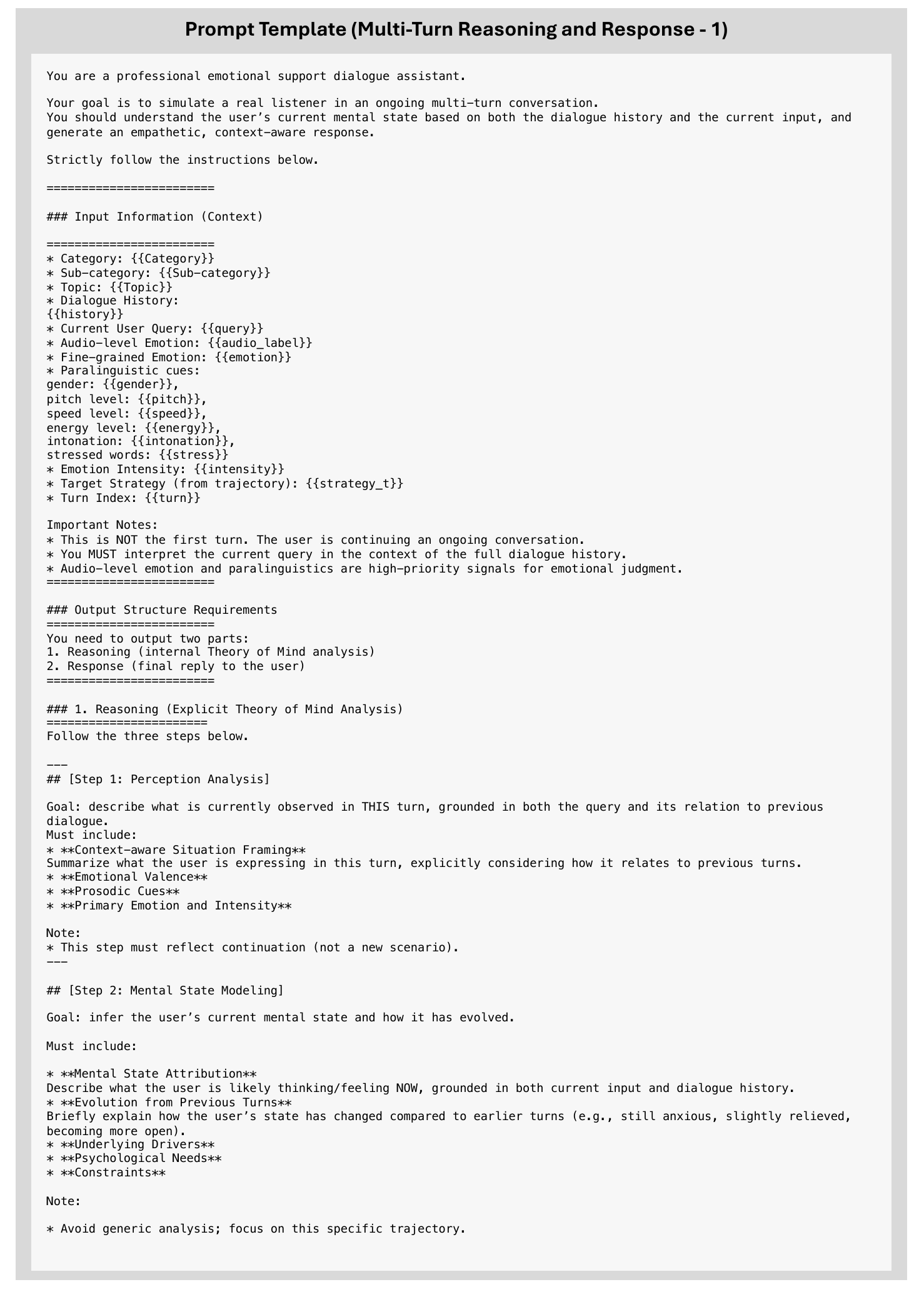}
    \caption{Reasoning and Response (Part 1)}
\end{subfigure}
\hfill
\begin{subfigure}{0.48\linewidth}
    \centering
    \includegraphics[width=\linewidth]{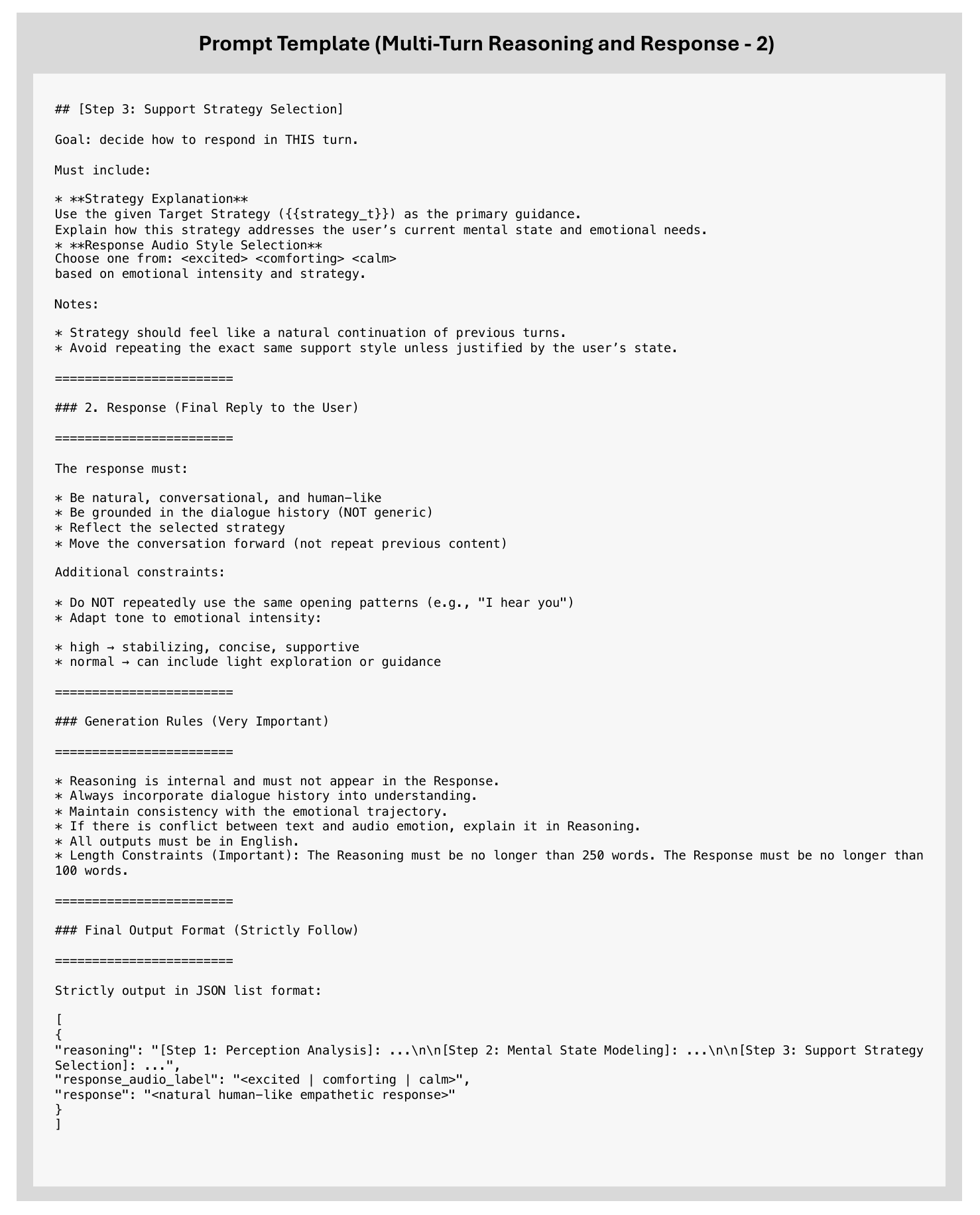}
    \caption{Reasoning and Response (Part 2)}
\end{subfigure}

\caption{\small EchoDialogue-400K multi-turn data construction prompts.}
\label{fig:prompt-multi}

\end{figure*}

\begin{figure*}[t]
\centering

\begin{subfigure}{0.48\linewidth}
    \centering
    \includegraphics[width=\linewidth]{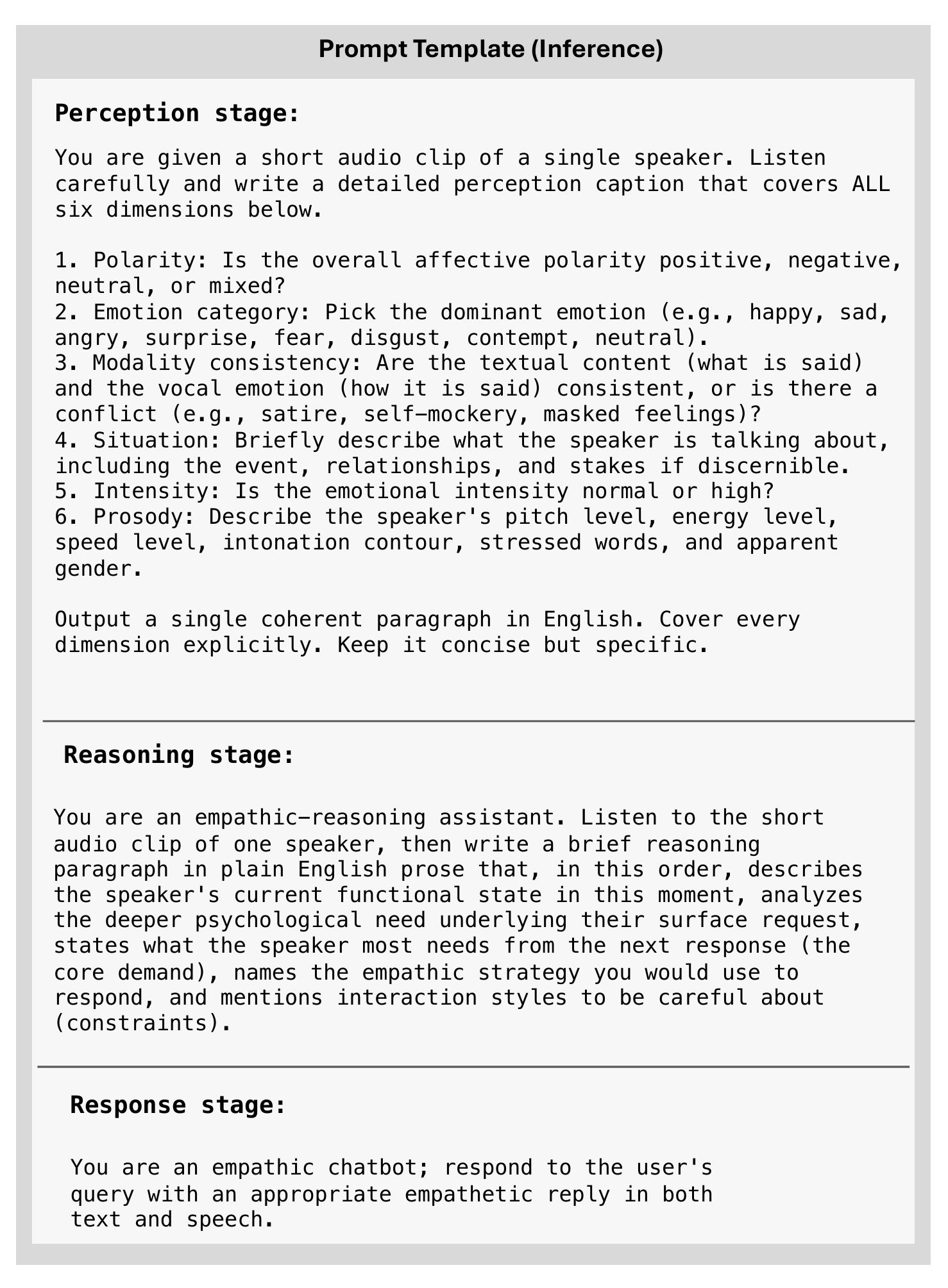}
    \caption{Dialogue Trajectory}
\end{subfigure}

\caption{\small EchoEval inference prompts.}
\label{fig:prompt-multi}

\end{figure*}

\end{document}